\documentclass[11pt,superscriptaddress,floatfix,showpacs]{revtex4}

\usepackage{graphicx}
\usepackage{dcolumn}
\usepackage{bm}

\usepackage[english]{babel}
\usepackage{amsmath}
\usepackage{amssymb}
\usepackage{amstext}
\usepackage{braket}
\usepackage{bbm}
\usepackage{relsize}
\usepackage{subfigure}
\usepackage{rotating} 

\newcommand{\ud}{\mathrm{d}}
\newcommand{\ue}{\mathrm{e}}

\newcommand{\vs}[1]{\ensuremath{\boldsymbol{#1}}}
\newcommand{\vect}[1]{\ensuremath{\bm{#1}}}

\DeclareMathOperator{\Tr}{Tr}

\begin{document}

\title[Modified spin-wave theory for frustrated bosons on the anisotropic triangular lattice]{Modified spin-wave theory with ordering vector optimization I: frustrated bosons on the spatially anisotropic triangular lattice}

\author{Philipp Hauke}
\email{Philipp.Hauke@icfo.es}
\affiliation{ICFO -- Institut de Ci\`encies Fot\`oniques, Av.\ Canal Ol\'impic s/n, E-08860 Castelldefels (Barcelona), Spain}
\affiliation{Max-Planck-Institut f\"ur Quantenoptik, Hans-Kopfermann-Str.\ 1, D-85748 Garching, Germany}
\author{Tommaso Roscilde}
\affiliation{Laboratoire de Physique, Ecole Normale Sup\'erieure de Lyon, 46 All\'ee d'Italie, F-69007 Lyon, France}
\author{Valentin Murg}
\affiliation{Max-Planck-Institut f\"ur Quantenoptik, Hans-Kopfermann-Str.\ 1, D-85748 Garching, Germany}
\author{J. Ignacio Cirac}
\affiliation{Max-Planck-Institut f\"ur Quantenoptik, Hans-Kopfermann-Str.\ 1, D-85748 Garching, Germany}
\author{Roman Schmied}
\affiliation{Max-Planck-Institut f\"ur Quantenoptik, Hans-Kopfermann-Str.\ 1, D-85748 Garching, Germany}

\date{\today}

\begin{abstract}

We investigate a system of frustrated hardcore bosons, modeled by an XY antiferromagnet on the spatially anisotropic triangular lattice, using Takahashi's modified spin-wave (MSW) theory.
In particular we implement ordering vector optimization on the ordered reference state of MSW theory, which 
leads to significant improvement of the theory and accounts for quantum corrections to the classically ordered state. 
The MSW results at zero temperature compare favorably to exact diagonalization (ED) and projected 
entangled-pair state (PEPS) calculations. The resulting zero-temperature phase diagram includes a 1D quasi-ordered 
phase, a 2D N\'{e}el ordered phase, and a 2D spiraling ordered phase. 
Strong indications coming from the ED and PEPS calculations, as well as from the
breakdown of MSW theory, suggest that the various ordered or quasi-ordered phases 
are separated by spin-liquid phases with short-range correlations, in analogy to what 
has been predicted for the Heisenberg model on the same lattice. 
Within MSW theory we also explore the finite-temperature phase diagram. 
In agreement with Berezinskii--Kosterlitz--Thouless (BKT) theory, we find that zero-temperature
long-range-ordered phases turn into quasi-ordered phases (up to a BKT transition temperature), while 
zero-temperature quasi-ordered phases become short-range correlated at finite temperature.
These results show that, despite its simplicity, modified spin-wave theory is very well suited
for describing ordered and quasi-ordered phases of frustrated XY spins (or, equivalently, 
of frustrated lattice bosons) both at zero and finite temperatures. 
While MSW theory, just as other theoretical methods, cannot describe spin-liquid phases, its breakdown provides a fast and reliable method for singling out Hamiltonians which may feature these intriguing quantum phases. We thus suggest a tool for guiding our search for interesting systems whose properties are necessarily studied with a physical quantum simulator instead of theoretical methods.

\end{abstract}

        
\maketitle


\begin{section}{Introduction}

Lattice models of strongly interacting bosons have recently been implemented experimentally thanks to  impressive developments with trapped ultra-cold atoms in optical lattice potentials \cite{Bloch2008,Lewenstein2007}.
A particularly appealing perspective in this arena is the study of 
strongly correlated lattice boson models with frustration, arising for instance from the coupling 
of bosons to a (artificial) magnetic field \cite{Jaksch2003, Sorensen2005, Polini2005}, 
or from a periodical shaking of the optical lattice \cite{Eckardt2009}. Frustration in the intersite hopping
amplitudes is formally equivalent to a description of the system in a rotating reference frame, 
which implies that the system is subject to the spontaneous appearance of vortices. Such
vortices can form ordered arrays (vortex crystals) coexisting with Bose condensation, 
which consequently takes place in a macroscopic wavefunction sustaining persisting 
circulating currents (see Ref.~\cite{Goldbaum2008} and references therein); 
or they can even disrupt condensation completely, 
and lead to a disordered insulating state \cite{Garcia-Ripoll2007}. 
Such disordered states are notoriously difficult to study theoretically.

In the particular limit of a very strong interparticle repulsion, frustrated bosonic 
models can be exactly mapped onto $S=1/2$ frustrated XY antiferromagnets \cite{Diep2004}.
In two dimensions these models are known to exhibit ground states with spiral order,  
representing the magnetic counterpart to the aforementioned Bose-condensed states 
with vortex arrays. In special circumstances the interplay between quantum 
fluctuations and frustration may lead to disordered spin-liquid states, 
which are in one-to-one correspondence with bosonic insulating phases.  
XY antiferromagnets can also be regarded as the limiting case of antiferromagnetic
Hamiltonians with planar anisotropy in the couplings, relevant to the description
of frustrated antiferromagnetic materials, and they can describe the physics 
of Cooper pairs in arrays of ultra-small Josephson junctions with magnetic
frustration \cite{Fazio2001}.
More recently we have proposed that frustrated XY antiferromagnets can be 
experimentally implemented by loading planarly trapped ions into an optical lattice \cite{Schmied2008}.

 From a theoretical point of view, bosonic frustration
in the presence of strong interparticle interactions on a lattice represents a very hard problem
in dimensions $d>1$, due to the lack of controlled perturbative expansions in the strongly
correlated regime; to the breakdown of semiclassical methods in the presence of 
strong quantum fluctuations enhanced by frustration; and to the appearance of a sign 
problem in quantum Monte Carlo simulations.
Hence the implementation of bosonic frustration in optical lattice experiments 
would represent a fundamental instance
of a \emph{useful} quantum simulation, possibly outperforming any classical computation (see \emph{e.g.\ }Ref.~\cite{Buluta2009} and references therein). 
Indeed, as mentioned above, solving frustrated bosonic models amounts
to solving a large class of frustrated antiferromagnets.
Fundamental steps have very recently been taken experimentally 
towards the implementation of artificial magnetic fields in cold atom experiments 
via Raman schemes \cite{Lin2009b}. Hence exciting progress in this field is expected in the near 
future. 
However, the difficulty of finding accurate theoretical descriptions of disordered quantum lattice models makes it hard to tell \emph{a priori} which systems will present such interesting phases in an experiment. 
The most attractive aspect of quantum simulators is their potential ability to emulate
model Hamiltonians whose phase diagram cannot be accurately predicted
with current theoretical approaches.
Therefore it would be highly desirable to dispose of a fast tool that can outline 
quantum-mechanical phase diagrams, point out disordered phases, and thus classify 
model Hamiltonians according to their potential interest for experimental quantum 
simulation. We propose that the methods presented here will serve this very purpose.

Planar systems of bosons in optical lattices, coupled to an artificial magnetic field, 
can be described by the Bose--Hubbard Hamiltonian
\begin{equation}
{\cal H}_{\mathrm{BH}} = \sum_{\braket{i,j}} \frac{t_{ij}}{2} \left(b_i^{\dagger} b_j + \text{h.c.}\right) + \frac{U}{2} \sum_i n_i (n_i - 1) 
\label{BH}
\end{equation}
where $b_i$, $b_i^{\dagger}$ are bosonic operators, $n_i = b_i^{\dagger} b_i$, and $\braket{i,j}$
represents pairs of nearest neighbor sites. The hopping amplitude 
$t_{ij} = - \tilde{t}_{ij} \exp\left({i A_{ij}}\right)$, where $\tilde{t}_{ij}\geq0$, is spatially modulated by the line integral of
the vector potential along the $\braket{i,j}$ bond, $A_{ij} = \int_{\bm{r}_j}^{\bm{r}_i} {\bm A}({\bm r})\cdot d\bm{l}$,
as well as by possible spatial anisotropies in the optical lattice (contained in the bare
hopping amplitudes $\tilde{t}_{ij} > 0$).
In the following we will focus on the limit of infinite repulsion $U\to\infty$ and half filling
$\langle n_i \rangle = 1/2$, under which the Bose--Hubbard model maps onto
 the $S=1/2$ XY Hamiltonian \cite{Schmied2008}:
\begin{equation}
 \label{HS}
  {\cal H_{\text{S}}}=
  \sum_{\braket{i,j}} t_{ij} \left(S_i^{\hspace{0.05cm}x} {\hspace{0.05cm}} S_j^{\hspace{0.05cm}x}
   + S_i^{\hspace{0.05cm}y} {\hspace{0.05cm}} S_j^{\hspace{0.05cm}y}\right)
\end{equation}
where $S_i^{\hspace{0.05cm}\alpha}$ is the $\alpha^{\text{th}}$ component of the $S=1/2$ spin operator acting on site $i$.  

In this work we focus on a triangular lattice with \emph {antiferromagnetic} nearest-neighbor interactions, which can be seen as a \emph{ferromagnetic} lattice ($\tilde{t}_{ij}>0$) with half a magnetic flux quantum threaded through each lattice plaquette. This magnetic flux can, for instance, be interpreted as flipping the signs of all hopping amplitudes of the bonds along the horizontal direction of the lattice (Fig.\ 1). In this gauge the hopping is transformed to $t_{i j} = \tilde{t}_{i j}$ for $r_i-r_j = \pm a\vect{\tau}_1$, where $a$ is the lattice spacing and $\vect{\tau}_1=\left(1,0\right)$, while $t_{i j}=-\tilde{t}_{i j}$ for the other (diagonal) nearest-neighbor bond directions. A canonical transformation rotating the spins in odd rows by an angle $\pi$ around the $z$ axis flips the signs of all diagonal bonds, leading to a model of a spatially anisotropic triangular XY antiferromagnet with all $t_{i j} > 0$. Such a model exhibits strong frustration on the triangular lattice. 
 In the following we specialize to the case in which the hopping amplitudes $t_{ij}$ take two values, $t_1$ and $t_2$
 depending on the orientation of the bond [parallel to $\tau_1$ or along the diagonals $\pm a \vect{\tau}_2$ and  $\pm a \left(\vect{\tau}_2-\vect{\tau}_1\right)$, with $\vect{\tau}_2=\left( 1/2, \sqrt{3}/2 \right)$ --
 see Fig.~\ref{fig:lattice}].
 This particular case is relevant for the physics of a number of systems: 
 it can describe neutral atoms trapped in triangular lattices formed with three lasers 
 intersecting at $120^\circ$ angles \cite{Becker2009}, and one of which has a different 
 intensity than the other two; another implementation of this model are neutral atoms in an \emph{isotropic} triangular lattice in the absence of artificial magnetic fields but with elliptical time-dependent forcing of the lattice \cite{Eckardt2009};
 furthermore, it is relevant for triangular Wigner crystals of ions trapped in the minima of an optical lattice \cite{Schmied2008}.
 
    \begin{figure}
        \centering
        \includegraphics[width=0.2\textwidth]{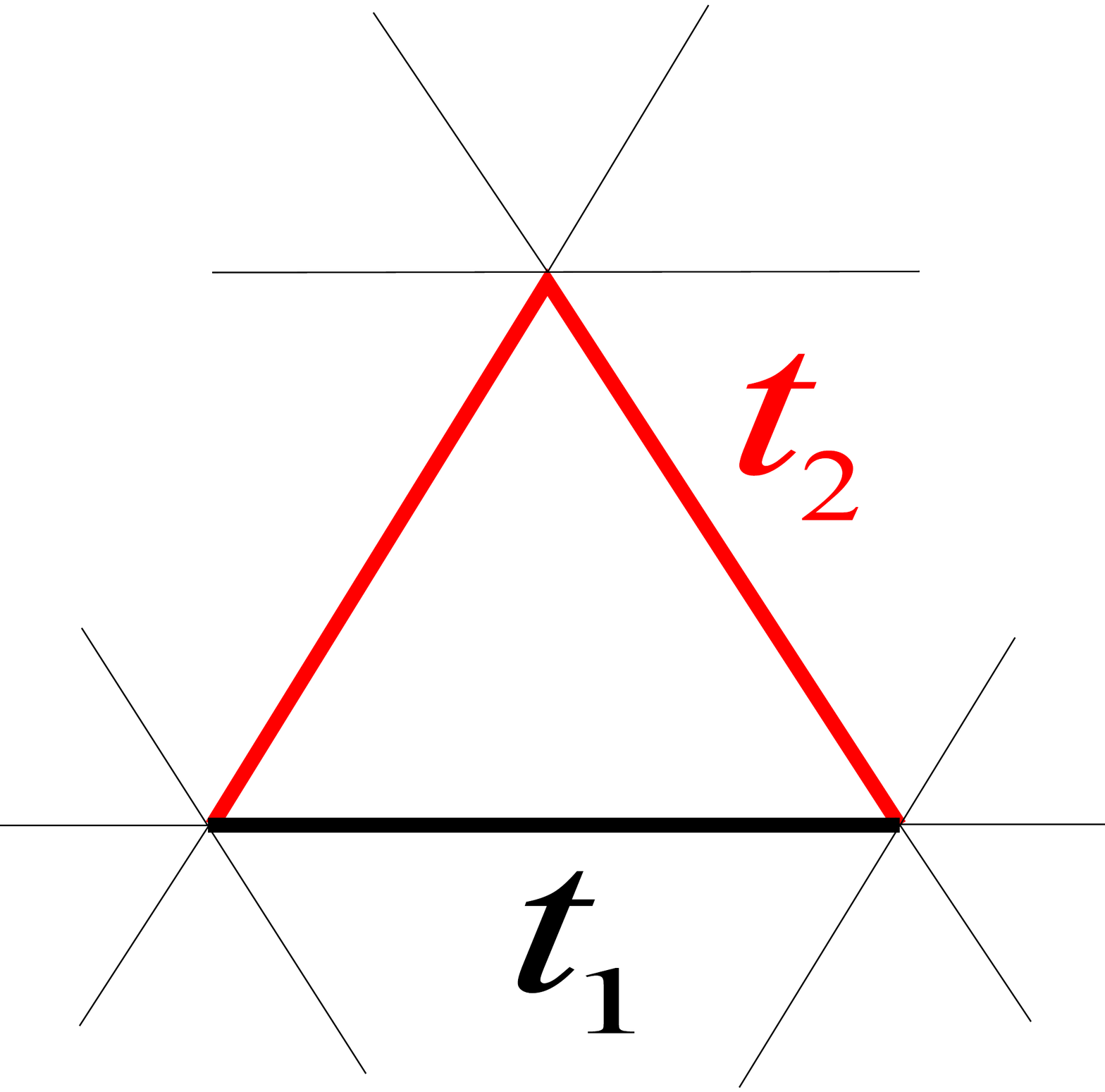}
        \caption{Spatially anisotropic triangular lattice. The lines denote interactions and the spins are located at the vertices.}
        \label{fig:lattice}
    \end{figure}

 The aim of this work is the determination of the ground-state and finite-temperature phase diagrams of the $S=1/2$ XY antiferromagnet (AF) - or, alternatively, 
 of frustrated half-filled hardcore bosons - on a spatially anisotropic 
triangular lattice (SATL) by means of spin-wave theory. In particular, we show that Takahashi's modified spin-wave 
(MSW) theory \cite{Takahashi1989} gives an adequate description 
of the main features of the ground-state and low-temperature phase diagram, while 
keeping the computational effort at a minimum. 
Spin-wave methods generally account for weak quantum fluctuations around the ordered state
corresponding to the ground state in the classical limit $S \to \infty$. 
We show how MSW theory can be extended to arbitrary reference states, 
whose ordering vector may be shifted with respect to the ground state in the classical limit \cite{Xu1991}.
Furthermore, we demonstrate how this procedure allows for a convenient calculation of the spin stiffness.
In the specific case of the $S=1/2$ XYAF on a SATL,  
the spin stiffness proves to be an effective tool supporting our search for spin-liquid phases. 
Supplementary results derived by projected entangled-pair states (PEPS) \cite{Verstraete2008} 
and exact diagonalizations (ED) using the Lanczos method allow us to validate the zero-temperature
results of the MSW method with ordering vector optimization. 
We moreover provide the finite-temperature phase diagram of the $S=1/2$ XYAS on a SATL, and find a region where the breakdown of MSW theory indicates a disordered state. Such a disordered phase is an ideal candidate for performing a \emph{useful} quantum simulation because theoretical tools for studying it adequately are currently lacking.
In an upcoming publication we will extend this method to the ground state phase diagrams of the Heisenberg SATL and the $J_1J_2J_3$ model \cite{Hauke2010UpcomingHeis}.

 For coherence with the theoretical technique used - spin-wave theory - and for a better comparison
 with existing results in the literature, the results of this paper will be generally 
 expressed in the language of spin physics, but guidance will be provided on how to 
 translate the magnetic observables into bosonic observables.  

This paper is organized as follows.
In section \ref{cha:msw} we introduce Takahashi's MSW formalism, supplemented with ordering-vector optimization, 
and provide a general method to calculate the spin stiffness. 
The rest of the work is dedicated to the investigation of the phase diagram of $S=1/2$ XYAF on a SATL.
Section \ref{cha:msw_triangNN} presents the ground state phase diagram of this model. 
In section \ref{cha:msw_triangFiniteT} we extend the phase diagram to finite temperatures and 
calculate Berezinskii--Kosterlitz--Thouless transition lines. Finally, in section \ref{cha:conclusion} we present our conclusions. 

\end{section}


\begin{section}{Modified spin-wave formalism\label{cha:msw}}

In the past MSW theory has been found to give a satisfactory qualitative account of the low-temperature properties of spin systems, even frustrated or disordered ones. 
In this section we review its formalism, mainly following Xu and Ting \cite{Xu1991} in the first steps,
but considering XY interactions rather than Heisenberg interactions (for the latter see also Ref.~\cite{Hauke2010UpcomingHeis}). This requires only minor modifications of the formulas, and
it is expected that the validity of the spin-wave approach is even better justified for XY interactions since in this case the influence of quantum fluctuations is reduced by the anisotropy in spin coupling.

Our aim is to determine the phase diagram of the Hamiltonian of Eq.~\eqref{HS}.
A fundamental assumption of spin-wave theory as applied to the XY model is that the 
ground state shows long-range order (LRO) with the spins classically lying in the $xy$-plane;
for a translationally invariant system, like the one under investigation, 
the ordered ground state is characterized by  
a well defined ordering vector $\vect{Q}$.
Under this assumption it is convenient to rotate the local reference system of each spin from 
$(x,y,z)$ to $(\eta,\zeta,\xi)$ so that the ground state 
in the local reference frame has all spins aligned in the same
direction. This amounts to the following transformation:

\begin{subequations}
\label{rot}
\begin{eqnarray}
S_i^{\,x}  &=& - \sin\left(\vect{Q}\cdot\vect{r}_i\right) S_i^{\,\eta} + \cos\left(\vect{Q}\cdot\vect{r}_i\right) S_i^{\,\zeta}\,,\\
    S_i^{\,y} &=& \phantom{-}\cos\left(\vect{Q}\cdot\vect{r}_i\right) S_i^{\,\eta} + \sin\left(\vect{Q}\cdot\vect{r}_i\right) S_i^{\,\zeta}\,,\\
    S_i^{\,z} &=& - S_i^{\,\xi}\,.
\end{eqnarray}
\end{subequations}
Then $S_i^{\,\zeta}$, which will be the quantization axis, lies parallel to the classical spin $\vect{S}_i=\left(\cos\left(\vect{Q}\cdot\vect{r}_i\right),\sin\left(\vect{Q}\cdot\vect{r}_i\right),0\right)$. The component $S_i^{\,\eta}$ lies perpendicular to it in the $xy$ plane, and $S_i^{\,\xi}$ is perpendicular to the $xy$ plane. 
Unlike in ordinary spin-wave theories we do not make any assumption on the ordering vector $\bm Q$.
In particular, it may well differ from the one exhibited
by the system in the classical limit ($\vect{Q}^{\rm cl}$). 

 Spin waves can be described by applying the Dyson--Maleev (DM) transformation \cite{Dyson1956,Maleev1957}, which maps the physical spins to interacting bosons, 
\begin{subequations}
\label{DM}
\begin{eqnarray}
    S_i^{\,-} &\to& \frac{1}{\sqrt{2 S}}\left(2 S - a_i^\dagger a_i\right)a_i\,,\\
    S_i^{\,+} &\to& \sqrt{2 S}\, a_i^\dagger,\\
    S_i^{\,\zeta} &\to& -S+a_i^\dagger a_i\,,
\end{eqnarray}
\end{subequations}
where $S_i^{\,\pm}\equiv S_i^{\,\xi}\pm i S_i^{\,\eta}$. 
The DM transformation is an exact mapping between spins and bosons as long as projectors are retained which keep the system 
in the physical subspace, \emph{i.e.\ }the subspace where at each site only $2 S$ DM bosons are present at most. 
It can be shown that these projectors have the form ${\cal P} = \mathbbm{1} + {\cal O}[n/(2S)]^3$ where $n$ is the DM
boson density \cite{Akhiezer1968}. Hence, under the assumption of diluteness of the DM boson gas,
$n/(2S)<1$ (in fact $\braket{n}=S$, see below), we can safely neglect the ${\cal P}$ projectors. 

 If the spin Hamiltonian under investigation is obtained as the hardcore limit of the Bose--Hubbard
 Hamiltonian Eq.~\eqref{BH}, it is important to distinguish the DM bosons from the physical 
 $b$-bosons from which the effective spin Hamiltonian originated. Indeed the DM bosons
 at a site $i$ quantify the deviation of the $i^{\text{th}}$ spin from the local direction in the 
 $xy$ plane set by the ordered structure with ordering vector $\bm Q$. On the other hand, 
 the physical bosons correspond to the spin deviations with respect to full alignement of 
 the spin along the $z$ axis. The particle-hole symmetry of the bosonic Hamiltonian,
 Eq.~\eqref{BH}, in the hardcore limit leads to half filling of the physical bosons, which
 accidentally coincides with the average filling imposed by the Takahashi's constraint
 on the DM bosons for $S=1/2$ (see next subsection). Yet all other properties are 
 in general quite different.

\subsection{Derivation of a mean-field Hamiltonian and Takahashi's constraint}
Applying Eqs.~\eqref{rot} and~\eqref{DM} to Eq.~\eqref{HS} one arrives at
the bosonic Hamiltonian

\begin{align}
\label{H4}
   {\cal H}&= S^2 \sum_{\braket{i,j}} t_{ij} \left[ 
    1 - \frac{1}{2S} \left(2 a_i^\dagger a_i + 2 a_j^\dagger a_j-a_i^\dagger a_j - a_i a_j^\dagger +a_i^\dagger a_j^\dagger+ a_i a_j\right)  \right.  \nonumber \\
    & 
    \left. + \frac{1}{(2S)^2} 
    \left( a_i^\dagger a_i a_j^\dagger a_j - a_i^\dagger a_j^\dagger a_j a_j - a_i^\dagger a_i a_i a_j^\dagger + a_i a_j^\dagger a_j a_j + a_i^\dagger a_i a_i a_j \right)
     + {\cal O}\left(\frac{n}{2S}\right)^3 \right]  \cos\left(\vect{Q}\cdot\vect{r}_{ij}\right)\,. 
\end{align}
Here we have dropped the terms with six boson operators, which are of order  ${\cal O}[n/(2S)^3]$ and are negligible for $n/(2S)<1$. 
Moreover the truncation of the Hamiltonian to this order is consistent with neglecting the effect of the projectors on the 
physical subspace which also amounts to discarding terms of order ${\cal O}[n/(2S)^3]$.

MSW theory relies on the minimization of the free energy. To this end we need the expectation value of the Hamiltonian Eq.~\eqref{H4}. 
Under the assumption that the ground state is a Gaussian state we make use of Wick's theorem 
\cite{Fetter1971} to decouple the boson--boson interaction terms, \emph{i.e.\ }the terms consisting of four boson operators. 
The expectation value $E\equiv\braket{\cal H}$ can then be written as
\begin{equation}
    E=\frac 1 2 \sum_{\braket{i,j}} t_{ij} \left\lbrace \left[S+\frac 1 2 - F\left( 0 \right) + F\left( \vect{r}_{ij} \right) \right]^2 + \left[S+\frac 1 2 - F\left( 0 \right) + G\left( \vect{r}_{ij} \right) \right]^2 \,\right\rbrace \cos\left(\vect{Q}\cdot\vect{r}_{ij}\right) \,.
\end{equation}
Here we have defined the correlators
\begin{subequations}
\begin{eqnarray}
    \braket{a_i^\dagger a_j}&=&F\left( \vect{r}_{ij} \right)-\frac 1 2 \delta_{ij}, \\
    \braket{a_i a_j}&=& \braket{a_i^\dagger a_j^\dagger} \,\, = \,\, G\left( \vect{r}_{ij} \right).
\end{eqnarray}
\end{subequations}
These correlators can be rewritten in terms of independent particles by first Fourier transforming $a_{\vect{k}}=\frac{1}{\sqrt{N}}\sum_i a_i\, \ue^{-i \vect{k}\cdot\vect{r}_i}$, where $N$ is the number of sites, and then applying a Bogoliubov transformation
\begin{subequations}
\label{Bogol}
\begin{eqnarray}
    \alpha_{\vect{k}\phantom{-}} &=& \phantom{-}\cosh\theta_{\vect{k}}\, a_{\vect{k}} - \sinh\theta_{\vect{k}} \, a_{-\vect{k}}^\dagger\,,\\
    \alpha_{-\vect{k}}^\dagger &=& -\sinh\theta_{\vect{k}} \, a_{\vect{k}} + \cosh\theta_{\vect{k}} \, a_{-\vect{k}}^\dagger\,.
\end{eqnarray}
\end{subequations}
Requiring that the Bogoliubov particles be non-interacting imposes $\braket{\alpha_{\vect{k}}\alpha_{\vect{k}'}}=\braket{\alpha^\dagger_{\vect{k}}\alpha^\dagger_{\vect{k}'}}=0$. This condition also removes the anti-Hermitian part of the Hamiltonian\label{page:DMnonhermitianremoved}. 

The correlators are now
\begin{subequations}
\begin{eqnarray}
    F\left( \vect{r} \right)&= & \frac 1 N \sum_{\vect{k}} \cosh\left(2 \theta_{\vect{k}}\right) e^{-i\vect{k}\cdot\vect{r}} \left(n_{\vect{k}}+\frac 1 2 \right),\\
    G\left( \vect{r} \right)&= & \frac 1 N \sum_{\vect{k}} \sinh\left(2 \theta_{\vect{k}}\right) e^{-i\vect{k}\cdot\vect{r}} \left(n_{\vect{k}}+\frac 1 2 \right),
\end{eqnarray}
\end{subequations}
with $n_{\vect{k}}=\braket{\alpha_{\vect{k}}^\dagger \alpha_{\vect{k}}}=1/\left(\exp\left(\omega_{\vect{k}}/T\right)-1\right)$ being the occupation number of Bogoliubov mode $\vect{k}$ at temperature $T$ (with the Boltzmann constant $k_\mathrm{B}$ set to unity). The dispersion relation $\omega_{\vect{k}}$ is determined self-consistently in the next section.

 So far we have essentially formulated a standard Hartree--Fock theory for the gas of interacting
 DM bosons. A very important modification to this theory, due to Takahashi \cite{Takahashi1989}, is the introduction 
 of the constraint of zero magnetization at each site,
\begin{equation}
    \label{constraint}
    \braket{S_i^{\,\zeta}}=-S+\braket{a_i^\dagger a_i}=-S-\frac 1 2 + F\left(0\right)=0.
\end{equation}
The implementation of this constraint amounts to effectively reducing the Hilbert space dimension available to the DM bosons by fixing their
average density to $S$.  For $S=1/2$ spins in a bipartite lattice one can in fact show a consequent reduction of the Hilbert space dimension from $\infty$ [as in linear spin-wave (LSW) theory] to $\frac{4}{\pi} \frac{2^N}{N}$ (as in MSW) which restores, up to logarithmic accuracy, the physical value of $2^N$ \cite{Dotsenko1994}.

Takahashi's constraint imposes that $\langle n \rangle /(2S) < 1$, in agreement with the kinematic constraint on the physical 
Hilbert space (even without explicit account of the projection operators on that space), and 
it guarantees the correctness of the truncations of high powers of $n/(2S)$ that we introduced above.   
Finally, if the Hamiltonian is $Z_2 \times U(1)$ symmetric because of a uniaxial 
anisotropy, as in the case of interest in this paper, 
one expects $\braket{S_i^{\,\zeta}}=0$. The constraint Eq.~\eqref{constraint} elegantly restores this 
reflection symmetry of the ground state with respect to the quantization axis.

\subsection{Derivation of the self-consistent equations\label{cha:msw_selfconsistentequations}}
The correct spin wave description is found by minimizing the free energy
$\mathcal{F}=E-T \mathcal{S}\,,$
where 
\begin{equation}
\label{entropy}
\mathcal{S}=\sum_{\vect{k}} \left[\left(n_{\vect{k}}+1\right) \ln\left(n_{\vect{k}}+1\right) - n_{\vect{k}} \ln n_{\vect{k}}\right]
\end{equation}
is the entropy of a set of harmonic oscillators. 
Minimizing with respect to $\theta_{\vect{k}}$ and $\omega_{\vect{k}}$ under the constraint of Eq.~\eqref{constraint} yields a set of self-consistent equations,
\begin{equation}
    \label{tanh2th}
    \tanh 2\theta_{\vect{k}}=\frac{A_{\vect{k}}}{B_{\vect{k}}}
\end{equation}
with
\begin{subequations}
    \label{AkBk}
\begin{eqnarray}
    \label{Ak}
    A_{\vect{k}} & = & -\frac 1 N \sum_{\braket{i,j}} t_{ij} \cos\left({\vect{Q}\cdot\vect{r}_{ij}}\right) G_{ij} \,\ue^{i\vect{k}\cdot\vect{r}_{ij}}\,,\\
    \label{Bk}
    B_{\vect{k}} &= & - \frac 1 N \sum_{\braket{i,j}} t_{ij} \cos\left({\vect{Q}\cdot\vect{r}_{ij}}\right) \left[ G_{ij} + F_{ij} \left(1-\ue^{i\vect{k}\cdot\vect{r}_{ij}}\right)\right] - \mu  \nonumber
\end{eqnarray}
\end{subequations}
where $\mu$ is the Lagrange multiplier for Eq.~\eqref{constraint} 
corresponding to the chemical potential for changing the total magnetization.

In Eqs.~\eqref{AkBk} we have abbreviated $F_{ij}=F\left(\vect{r}_{ij}\right)$, and $G_{ij}=G\left(\vect{r}_{ij}\right)$. Note that in the classical limit $S\to\infty$ one gets $G_{ij},F_{ij}\approx S$ and Eqs.~\eqref{AkBk} become analogous to their LSW counterparts.
The spin-wave spectrum reads
\begin{equation}
\label{disp}
\omega_{\vect{k}}=\sqrt{B_{\vect{k}}^2-A_{\vect{k}}^2}\,.
\end{equation}
Inserting Eq.~\eqref{AkBk} into Eq.~\eqref{disp} shows that a finite $\mu$ entails a gap at $\vect{k}=0$. This is to be seen in contrast to LSW theory where the spectrum always has a gapless Goldstone mode at $\vect{k}=0$.
The correlators at the minimum take the form
\begin{subequations}
\label{FGfiniteT}
\begin{eqnarray}
    F_{ij}&=&\frac{1}{N} \sum_{\vect{k}} \frac{B_{\vect{k}}} {\omega_{\vect{k}}}\cos\left(\vect{k}\cdot\vect{r}_{ij}\right) \left(n_{\vect{k}}+\frac{1}{2}\right)\label{Fij}\,,\\
    G_{ij}&=&\frac{1}{N} \sum_{\vect{k}} \frac{A_{\vect{k}}} {\omega_{\vect{k}}}\cos\left(\vect{k}\cdot\vect{r}_{ij}\right) \left(n_{\vect{k}}+\frac{1}{2}\right)\label{Gij}\,.
\end{eqnarray}
\end{subequations}

At $T=0$ where $n_{\vect{k}}=0\,\, \forall \vect{k}\neq 0$, one finds that $\mu$ vanishes\label{page:vanishinggap}. This implies also the disappearance of the gap at $\vect{k}=0$ that may exist for finite temperature. A vanishing gap is a necessary requirement for the appearance
of the Goldstone mode associated with magnetic LRO.
It also enables Bose condensation of the DM bosons in the $\vect{k}=0$ mode. 
 Separating out the contribution of the zero mode, 
$\braket{a_{\vect{k}=0}^{\dagger} a_{\vect{k}=0}}/N=\braket{a_{\vect{k}=0} a_{\vect{k}=0}}/N\equiv M_0$
(corresponding to the order parameter measuring the total spiraling magnetization in the quantization axis directions given by the ordering vector $\bm{Q}$), one arrives at the zero-temperature equations
\begin{subequations}
\label{FG}
\begin{eqnarray}
    F_{ij}&=&M_0 + \frac 1 {2 N} \sum_{\vect{k}\neq 0}\!\! ~\frac{B_{\vect{k}}} {\omega_{\vect{k}}}\cos\left(\vect{k}\cdot\vect{r}_{ij}\right)\label{FijS}\,,\\
    G_{ij}&=&M_0 + \frac 1 {2 N} \sum_{\vect{k}\neq 0}\!\! ~\frac{A_{\vect{k}}} {\omega_{\vect{k}}}\cos\left(\vect{k}\cdot\vect{r}_{ij}\right)\label{GijS}\,,
\end{eqnarray}
\end{subequations}
and the constraint Eq.~\eqref{constraint} becomes
\begin{equation}
    \label{constr2}
    S+\frac 1 2 = M_0 + \frac 1 {2 N} \sum_{\vect{k}\neq 0} \!\! ~\frac{B_{\vect{k}}} {\omega_{\vect{k}}}.
\end{equation}
As mentioned above, the occupation of the zero mode $M_0$ corresponds to a Bose--Einstein condensate of the DM bosons in the minimum of the dispersion relation.
This condensate is depleted by interactions of the DM bosons. The larger this depletion, the more DM bosons reside at momenta different from zero, thus decreasing magnetic LRO.

\subsection{Optimization of the ordering vector}
It is not \emph{a priori} clear that the classical ordering vector $\vect{Q}^{\mathrm{cl}}$ correctly describes the LRO in the quantum system. 
To account for the competition between states with LRO at different ordering vectors $\vect{Q}$ we extend the MSW procedure 
by optimizing the free energy with respect to the ordering vector $\vect{Q}$.
This procedure, first introduced in Ref.~\cite{Xu1991}, significantly improves
the predictions of MSW theory.
 It amounts to finding the best ordered reference state with in-plane ordering vector  $\vect{Q}$
 (spiral state) whose free energy  is minimized 
 not at the classical level, but including the effect of
 quantum fluctuations self-consistently within MSW theory.

The minimization of $\mathcal{F}$ with respect to $Q_x$ and $Q_y$ yields two additional equations which must be added to the set of self-consistent equations, 
\begin{subequations}
\label{Qs}
\begin{equation}
    \label{Qx}
    \frac\partial{\partial Q_x} \mathcal{F}=-\frac 1 2 \sum_{\braket{i,j}} t_{ij} \sin\left(\vect{Q}\cdot\vect{r}_{ij}\right)r_{ij}^x\left[F_{ij}^2+G_{ij}^2\right] = 0 \,,
\end{equation}
\begin{equation}
    \label{Qy}
    \frac\partial{\partial Q_y} \mathcal{F}=-\frac 1 2 \sum_{\braket{i,j}} t_{ij} \sin\left(\vect{Q}\cdot\vect{r}_{ij}\right)r_{ij}^y\left[F_{ij}^2+G_{ij}^2\right]=0\,.
\end{equation}
\end{subequations}

The values of $F_{ij}$ and $G_{ij}$ can now be calculated by solving self-consistently Eq.~\eqref{Qs} together with 
Eqs.~(\ref{constraint},~\ref{tanh2th}--\ref{FGfiniteT}). At zero temperature, Eq.~\eqref{constraint} and Eq.~\eqref{FGfiniteT} have to be replaced by Eq.~\eqref{constr2} and Eq.~\eqref{FG}, respectively.
Through Wick's theorem the knowledge of the quantities $F_{ij}$ and $G_{ij}$ 
allows for the computation of the expectation value of any observable.

\subsection{Spin stiffness\label{cha:gausscurvtriang}}

The optimization of the ordering vector allows for a straightforward calculation of the spin stiffness. 
This additional information, complementary to the order parameter, helps us in identifying candidate regions for spin-liquid behavior. 

The MSW Ansatz always returns only a single one of all the possible ordering vectors $\vect{Q}^0$ as the optimal ordering vector. However, if the true ground state is only short-range ordered, we might expect the $\vect{Q}$-minimum to be relatively shallow, and that a slight change of the ordering vector barely affects the free energy $\mathcal{F}$.
This means that the order is not very stable against twists of the spin configuration.
We quantify the curvature of the minimum by the spin stiffness tensor 
\begin{equation}
\rho_{\alpha\beta}=\frac 1 N \left. \frac{\ud^2 \mathcal{F}}{\ud Q_\alpha \ud Q_\beta}\right|_{\vect{Q}=\vect{Q}^0}\,,
\end{equation}
evaluated at the optimized ordering vector $\vect{Q^0}$. In particular we will determine the \emph{parallel spin stiffness}
$
\rho_{\|}\equiv\frac{1}{2}\Tr \rho = \frac{1}{2}\left(\rho_{xx}+\rho_{yy}\right)
$
and the \emph{Gaussian spin stiffness}
\begin{equation}
\label{Upsilon}
\Upsilon=\det \rho\,.
\end{equation}

The spin stiffness gives a measure of how stiff magnetic LRO order is with respect to distortions of the ordering vector and it provides a fundamental self-consistency check of our approach. In fact, finding a small spin stiffness casts doubt on the reliability of the spin-wave approach in describing such a strongly fluctuating state, and hence suggests that the true ground state might be quantum disordered. 
 
Since a change in $\vect{Q}$ affects the correlators $F_{ij}$ and $G_{ij}$, we must compute $\Upsilon$ self-consistently. After finding the optimal $\vect{Q}^0$ by the self-consistent procedure described in the previous sections, we calculate $\frac{1}{N} \mathcal{F}\left(Q_x,Q_y\right)$ self-consistently for several ordering vectors $\vect{Q}=\vect{Q}^0+\Delta\vect{Q}$ and fit a quadratic form to the results. Since the minimum in the free energy can be very shallow, this procedure can be somewhat affected by numerical noise. 
As an approximation to the true spin stiffness, the \emph{partial spin stiffness} $\rho_{\alpha\beta}^{\mathrm{partial}}$ can be computed via the partial derivatives, \emph{i.e.\ }without recalculating the self-consistent equations. It reads
\begin{eqnarray}
    \label{gcpartial}
    \rho_{\alpha\beta}^{\mathrm{partial}}&\equiv&\frac 1 N \frac{\partial^2}{\partial Q_{\alpha}\partial Q_{\beta}} \mathcal{F}\\
    &=&-\frac 1 {2N} \sum_{\braket{i,j}} t_{ij} \cos\left(\vect{Q}\cdot\vect{r}_{ij}\right)r_{ij}^{\alpha}r_{ij}^{\beta}\left[F_{ij}^2+G_{ij}^2\right] \nonumber \,.
\end{eqnarray}
We define $\Upsilon^{\mathrm{partial}}$ analogously to $\Upsilon$ [Eq.~\eqref{Upsilon}] as the determinant of the partial spin-stiffness tensor.
The system can lower its energy by adjusting $F_{ij}$ and $G_{ij}$ to the new ordering vector, and therefore $\Upsilon$ is always smaller than $\Upsilon^{\mathrm{partial}}$.
We will see later that in some cases the partial Gaussian spin stiffness $\Upsilon^{\mathrm{partial}}$ gives a good estimate of the real Gaussian spin stiffness $\Upsilon$, 
but there are cases where it is considerably larger.

In the following we present the zero-temperature phase diagram of the anisotropic triangular lattice.
This paradigmatic example will show that the spin stiffness is an important quantity that provides a deeper 
insight into the order properties of the system.

\subsection{From spins to bosons \label{cha:spinboson}} 

As we mentioned in the introduction, in the $S=1/2$ case the spin Hamiltonian is equivalent
to that of infinitely repulsive bosons at half filling with frustrated hoppings. Hence it is important to match spin observables with their bosonic counterparts. Following the Dyson--Maleev or the Holstein--Primakoff
transformation, spin operator and physical (hardcore) $b$ bosons obey the relationship: 
$\tilde{S}_i^+ = b^{\dagger}_i$, $\tilde{S}_i^- = b_i$, $\tilde{S}_i^z=S_i^z = b^{\dagger}_i b_i - 1/2$ \footnote{Here the operator $\tilde{S}_i^+$ ($\tilde{S}_i^-$) raises (lowers) the spin with respect to the $z$ axis, in contrast to $S_i^+$ and $S_i^-$ (as defined in section~\ref{cha:msw}) which act with respect to the quantization axis $\zeta$ of the twisted coordinate system, Eq.~\eqref{rot}.}. 
Here the $b$ operators obey \emph{anticommutation} rules on site, $\{b_i, b_i^{\dagger} \} = 1$.
A non-zero magnetic order parameter $M_0$ implies the appearance of off-diagonal LRO in the bosonic one-body density matrix, $\braket{b_i^\dagger b_j}\stackrel{\left|\vect{r}_{ij}\right|\to\infty}{\longrightarrow}M_0^2\,\cos\vect{Q}\cdot\vect{r}_{ij}$. 
The ordering vector $\bm Q$ corresponds to the finite momentum at which condensation occurs. 
The condensed state in the spiral phase is characterized by a pattern of persistent currents forming 
a crystal of vortices, whose geometrically correlated structure is captured by the spin chirality (see below). 
Finally the parallel spin stiffness corresponds to the superfluid density of the bosons, $\rho_s=\rho_{\|}/S$.  

\end{section}


\begin{section}{\label{cha:msw_triangNN}Ground state phase diagram of the anisotropic triangular lattice}

In this section we compute the ground-state phase diagram of the spatially anisotropic triangular lattice (SATL) with nearest-neighbor (NN) XY interactions. 
We consider a wide range of $\alpha\equiv t_{2}/t_{1}$, where $t_1$ denotes the bond strengths along the chains and $t_2$ the bond strengths 
along the diagonals (black and red bonds, respectively, in Fig.~\ref{fig:lattice}). The parameter $\alpha$ interpolates between an ensemble of decoupled one-dimensional chains at $\alpha=0$, the isotropic triangular lattice at $\alpha=1$, and the square lattice for $\alpha \to \infty$.
 
If we assume the spins to behave classically, the 2D-N\'{e}el order, present for $\alpha \geq 2$, starts to continuously
deform into spiral order at $\alpha \leq 2$ [compare Fig.~\ref{fig:phasediagtriang}~(a)]. 
The spiral phase extends down to $\alpha=0$ where the chains decouple.
In a previous publication we presented the quantum mechanical phase diagram as predicted by projected entangled-pair states (PEPS) calculations \cite{Schmied2008}; it is reproduced for convenience in Fig.~\ref{fig:phasediagtriang}~(b).
According to this, both the square lattice limit ($\alpha\to\infty$) and the most frustrated case, the isotropic triangular lattice ($\alpha=1$), display magnetic LRO. In the limit of uncoupled chains ($\alpha=0$) the system displays quasi-LRO with algebraically decaying correlations. 
However, similarly to what has been found in the Heisenberg model \cite{Yunoki2006}, the system seems to feature spin-liquid phases with exponentially decaying 
correlations between different types of order or quasi-order. In Appendix A we provide a further spectral feature, coming from the exact diagonalization
of a small cluster, which is consistent with the observation coming from PEPS calculations.
Further distinct features of the quantum model are that the transition between 2D-N\'{e}el and spiral order is shifted by 
quantum fluctuations to considerably smaller values of $\alpha$, and that the quasi-ordered 1D-like state extends over 
a whole region of finite $\alpha$ in the phase diagram. 

    \begin{figure}
        \centering
        \includegraphics[width=0.6\textwidth]{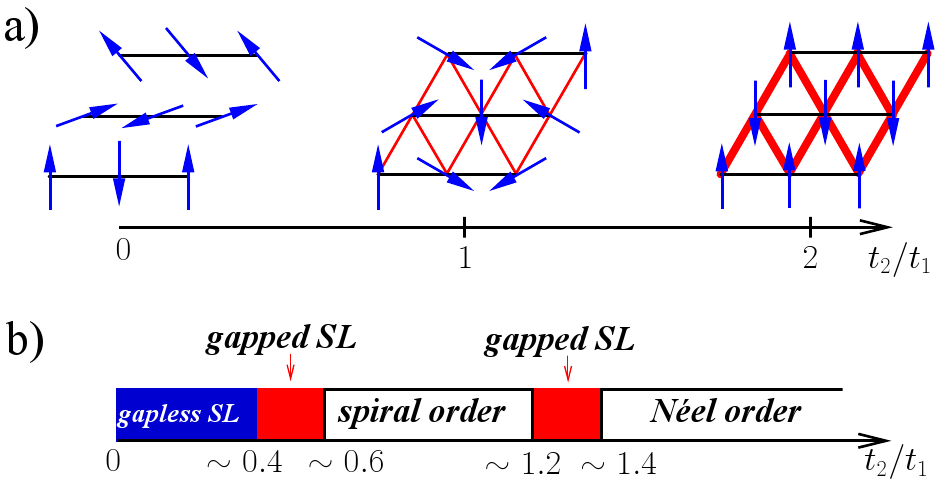}
        \caption{(a) Classical ground-state phase diagram of the anisotropic triangular lattice with sketches of the 1D state at $\alpha=0$, the spiral state at $\alpha=1$ and the 2D-N\'{e}el state for $\alpha\geq 2$. (b) Quantum mechanical phase diagram from Ref.~\cite{Schmied2008}. SL is short for spin liquid. }
        \label{fig:phasediagtriang}
    \end{figure}
In the following we compare predictions of MSW theory with these PEPS results and exact diagonalizations (ED). This will allow us to validate the reliability of the MSW method. 
The following system geometries are considered for the three different methods:
\begin{itemize}
\item
PEPS: a rhombic lattice of $20\times 20=400$ spins with open boundary conditions and bond dimension $D=2$. PEPS is a powerful numerical tool which goes beyond mean-field theory, but which, for small bond dimension $D$, only partially accounts for the entanglement in the ground state.
This limitation becomes particularly serious close to quantum phase transitions. 
However, in Ref.~\cite{Schmied2008} it was demonstrated that $D=2$ is already 
accurate enough to effectively capture the physics of the system.
\item
ED: Lanczos diagonalization of clusters of 24 and 30 spins (the latter is shown in Fig.~\ref{fig:systemsexactdiag}), 
again with open boundary conditions (necessary in order to allow for the accommodation of arbitrary ordering vectors). 
\item
MSW: rhombic lattices of $32\times 32=1024$ and $64\times 64=4096$ spins and in the infinite-lattice (thermodynamic) limit, under periodic boundary conditions. We find that at these lattice sizes all quantities have essentially reached the infinite lattice limit in most of parameter space. As it can be expected, the deviations from the thermodynamic limit are sizable only in the one-dimensional limit and at critical points. The thermodynamic limit is computed by replacing finite sums over the first Brillouin zone with integrals.

\end{itemize}
    \begin{figure}
        \centering
        \includegraphics[width=0.15\textwidth]{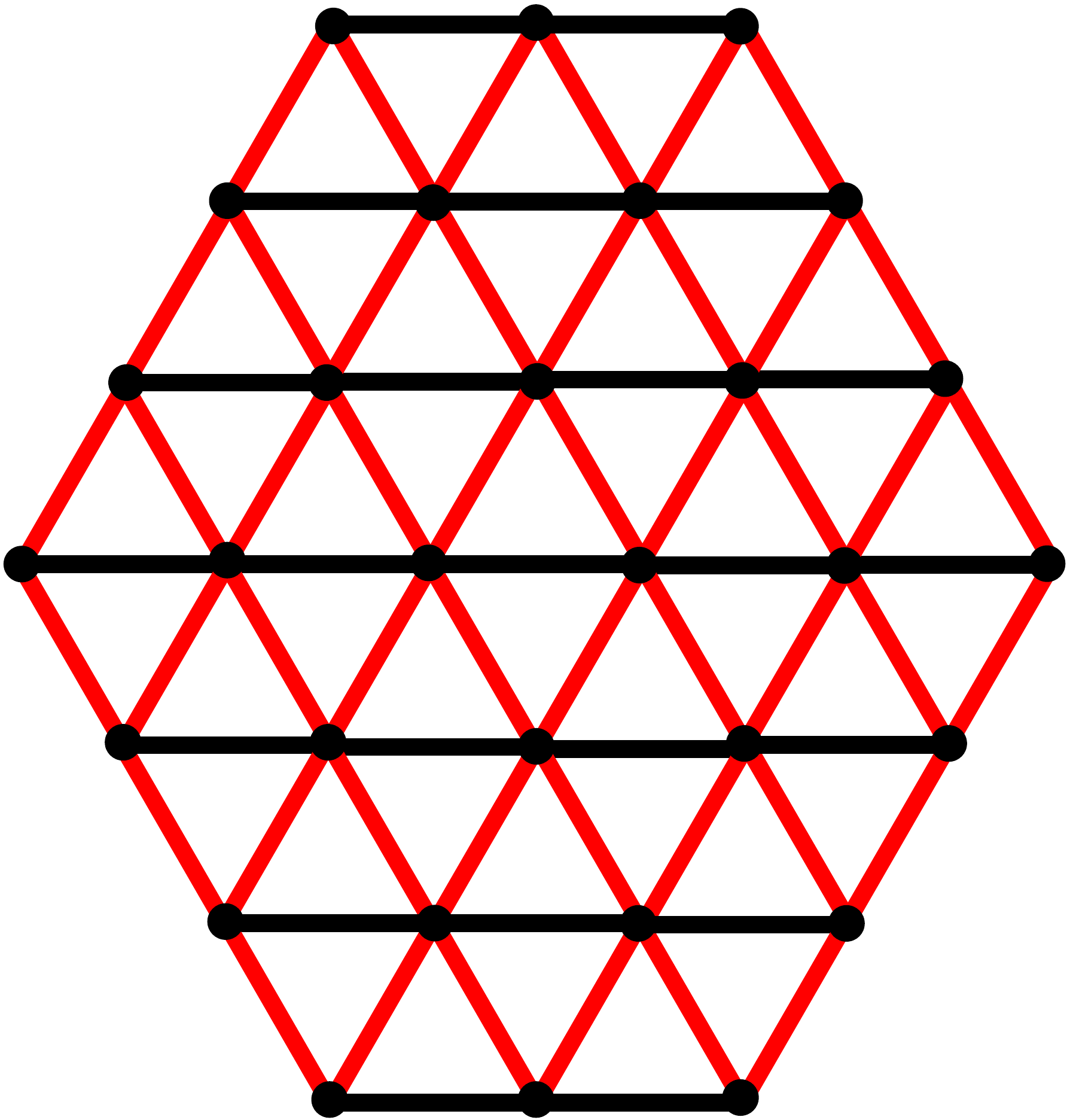}
        \caption{Cluster of 30 spins for which we carried out exact diagonalizations (ED). The 24-spin system is equivalent, only with the top and bottom rows removed. The clusters are chosen for largest symmetry with respect to reflection on the axes and for a ratio of $t_2$- (red) to $t_1$-bonds (black) closest to 2.
        }
        \label{fig:systemsexactdiag}
    \end{figure}

In the triangular lattice with nearest-neighbor interactions one finds in the MSW formalism that Eq.~\eqref{Qy} gives $Q_y=0$ 
and from Eq.~\eqref{Qx} we obtain the formula
\begin{equation}
    \label{Qx2}
    Q_x=2\arccos\left[-\frac \alpha 2 \frac{F_{\vs{\tau}_2}^2+G_{\vs{\tau}_2}^2}{F_{\vs{\tau}_1}^2+G_{\vs{\tau}_1}^2}\right]\,.
\end{equation}
Here $\vs{\tau}_1=\left(1,0\right)$ and $\vs{\tau}_2=\left(1/2,\sqrt{3}/2\right)$ are the primitive lattice vectors (here and in the rest of the work we take the lattice spacing $a$ equal to unity).
For $F_{ij}=G_{ij}= S$, attained when $S\to \infty$, this reduces to the ordering vector of the LSW theory $\left(Q_x^{\text{cl}},Q_y^{\text{cl}}\right)=\left(2 \arccos\left(-\alpha/2\right),0\right)$ which coincides with the classical ordering vector. 

\subsection{Breakdown regions for MSW theory}

As a first step in our analysis we investigate the parameter regions where LRO is to be expected, and the locations where MSW theory ceases to be applicable. 
To this end we first investigate if there appear imaginary modes in the dispersion relation, which would indicate instabilities.
Afterwards we study the order parameter $M_0$ and the spin stiffness.

\subsubsection[Imaginary frequencies and breakdown of convergence]{Imaginary frequencies and breakdown of convergence}
 
  Convergence in the self-consistent equations of MSW theory with ordering vector optimization, 
  Eqs.~(\ref{tanh2th}--\ref{disp},~\ref{FG},~\ref{constr2},~\ref{Qx2}), 
  cannot be achieved in selected regions of the ground-state phase diagram, namely
  for $\alpha \lesssim 0.18$ and for $1.35 \lesssim \alpha \lesssim 1.66$, 
  as summarized in Fig. \ref{fig:TLXY_imaginaryModes}.
  (Interestingly, convergence is restored in the pure 1D limit, $\alpha=0$, 
  for which the theory formulates surprisingly good predictions.)  
  This breakdown of convergence corresponds to 
  the appearance of an imaginary part in the spin-wave frequencies, Eq.~\eqref{disp}, 
  signaling an instability of the ordered ground state.  
  The breakdown of a self-consistent description of the system in terms of an ordered ground state is strongly suggestive
  of the presence of a quantum-disordered ground state in the exact behavior of the system. 
  Hence, one can interpret these parameter regions as candidates for the spin-liquid phases predicted by 
  PEPS calculations \cite{Schmied2008} [compare Fig.~\ref{fig:phasediagtriang}~(b)].
  
   \begin{figure}[h]
        \centering
        \includegraphics[width=0.6\textwidth]{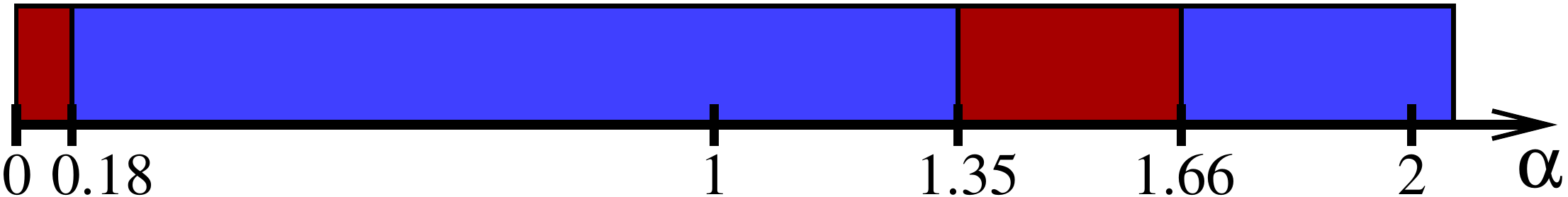}
        \caption{Regions in the phase diagram in which imaginary modes appear and convergence of the MSW equations breaks down (red).}        
        \label{fig:TLXY_imaginaryModes}
    \end{figure}

\subsubsection[Order parameter and spin stiffness]{Order parameter and spin stiffness}

A fundamental indication for the validity of spin-wave theories is generally given by the order parameter $M_0$ (Fig.~\ref{fig:M_XY_triang}) and the spin stiffness (Fig.~\ref{fig:gc_XY_triang}). 
The influence of quantum fluctuations is strong where they are small, and the primary assumption that the system can be described by a semi-classical spin-wave state begins to falter in such a case. 
Since the MSW formalism only takes quantum fluctuations partially into account,  a small order parameter and/ or spin stiffness also suggests that the true quantum ground state could be completely disordered.

    \begin{figure}
        \centering
        \includegraphics[width=0.6\textwidth]{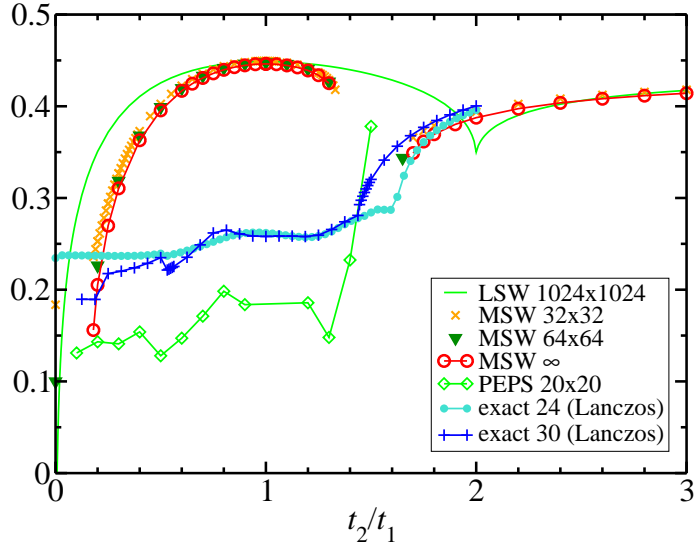}
        \caption{Comparison of the MSW order parameter $M_0$ [as defined before Eq.~\eqref{FG}] for different system sizes (numbers behind the labels). Also included is $\sqrt{S\left(\vect{Q}\right)/N}$ [Eq.~\eqref{Sk}] for ED (Lanczos) and PEPS computations, and the staggered magnetization $M$ of LSW theory. A large value indicates strong LRO, with the theoretical maximum being $0.5$.
        }
        \label{fig:M_XY_triang}
    \end{figure}
\begin{figure}
        \centering
        \includegraphics[width=0.6\textwidth]{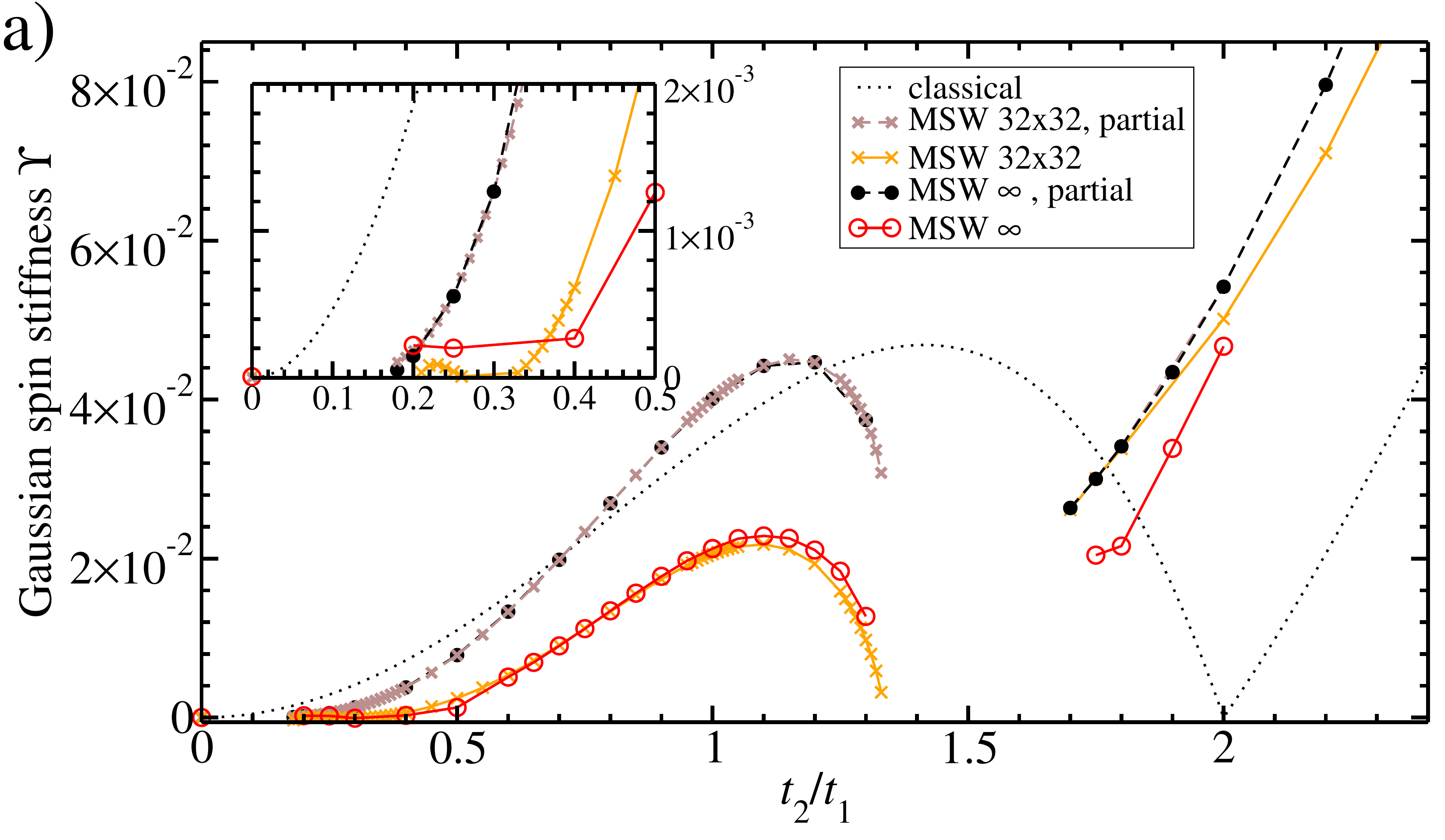}
        \includegraphics[width=0.6\textwidth]{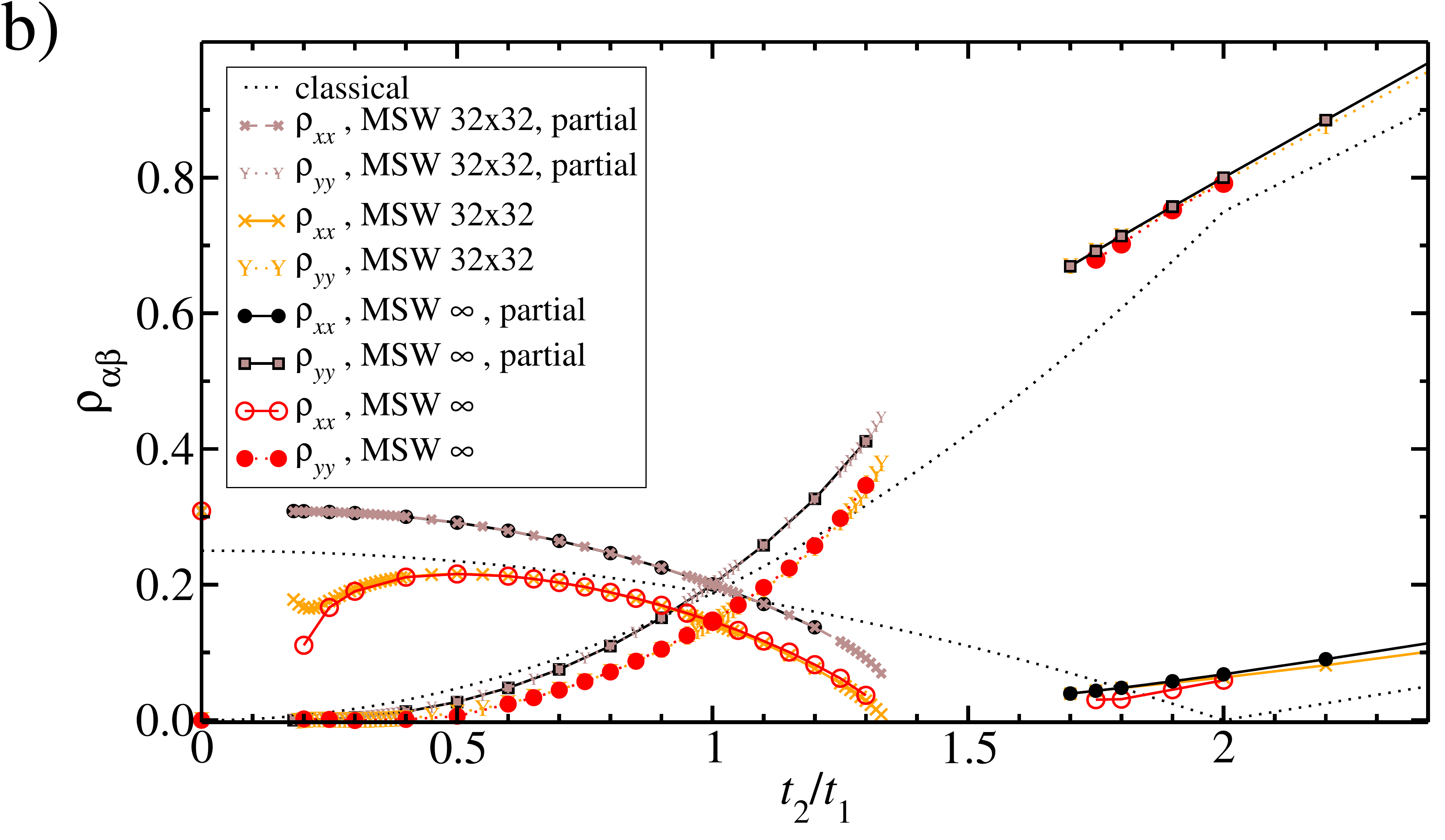}
        \caption[Gaussian spin stiffness]{(a) Gaussian spin stiffness $\Upsilon$; (b) components of the spin stiffness tensor. The mixed second derivative $\rho_{xy}$ vanishes for symmetry reasons. The numbers behind the labels of the graphs give the considered system size. The inset in (a) is a close-up of the region of small $\alpha$. The curves labeled `partial' were obtained by Eq.~\eqref{gcpartial}.
        }
        \label{fig:gc_XY_triang}
\end{figure}

Interestingly in the regions of largest spatial isotropy of the interactions, \emph{i.e.\ }around $\alpha=1$ (isotropic triangular lattice) and at large $\alpha$ (isotropic square lattice), the order parameter $M_0$ coincides with that of LSW theory. As we will see later, at the points $\alpha=0,1,\infty$ the ordering vector found with the present MSW approach also exactly matches the classical one, due to symmetry. 

In the square lattice limit $\alpha\to\infty$, the order parameter attains the value $M_0=0.435$ in the thermodynamic limit, which is very close to $M=0.437$ as extrapolated from quantum Monte Carlo calculations \cite{Sandvik1999}.
For the spin stiffness Ref.~\cite{Sandvik1999} obtained $\rho_{\|}/\alpha=0.270$; 
the MSW method returns the only slightly larger value $\rho_{\|}/\alpha=0.272$.
It appears that in this special case the main quantum corrections are correctly captured by MSW theory. The large values of the 
order parameter (around 87\% of the theoretical maximum) and of the spin stiffness support the assumption that the classical picture remains essentially valid in the large-$\alpha$ limit.

The loss of LRO in the 1D-limit ($\alpha\to0$) is reflected in the breakdown of the order parameter $M_0$, which occurs at a finite value of the inter-chain coupling, $\alpha\approx 0.18$ (note that within LSW theory the order parameter vanishes only for $\alpha\to0$). 
This coincides with the appearance of imaginary spin-wave energies as discussed in the previous section. 
At small but finite $\alpha$ the spin stiffness $\rho_{yy}$ essentially vanishes, which is characteristic of a 1D-like state that consists of effectively decoupled chains. This suggests that the physics becomes basically independent of the $y$-component of $\vect{Q}$ for $\alpha\lesssim 0.35$. 

A single XY chain can be solved exactly by Bethe-Ansatz equations, and by use of twisted boundary conditions one can obtain the exact solution for the spin stiffness $\rho_{xx}=1/\pi\approx 0.318$ \cite{Shastry1990}. \label{page:rhoxx_XYchain} 
Our MSW result of $\rho_{xx}\approx 0.308$ lies surprisingly close.
For one-dimensional models it is known that a non-zero spin stiffness is accompanied by quasi long-range correlations with power-law decay.
The critical nature of the state in the 1D-like phase reflects itself also in the fact that
finite-size effects play an important role. 

The MSW order parameter can be compared with results from exact diagonalization (ED) and PEPS calculations. 
In both cases the 
Fourier transform of the spin--spin correlations, the static structure factor
\begin{equation}
    \label{Sk}
    S\left(\vect{k}\right)= \frac{1}{N} \sum_{i,j}e^{-i \vect{k} \cdot(\vect{r}_{i}-\vect{r}_j)}  \braket{S_{i}^x S_{j}^x+S_{i}^y S_{j}^y} \,,
\end{equation}
allows to extract an ordering vector $\vect{Q}$ which maximizes $S\left(\vect{k}\right)$, as well as the order 
parameter $M$ which is defined as $M=\sqrt{S\left(\vect{Q}\right)/N}$ in the thermodynamic limit.
In Fig.~\ref{fig:M_XY_triang} we compare $\sqrt{S\left(\vect{Q}\right)/N}$ for ED of systems of 30 and 24 spins, 
and with PEPS calculations on a $20\times 20$ lattice. 

The comparison of ED and MSW results shows that MSW is quantitatively reliable in the N\'eel phase for $\alpha \gtrsim 1.66$. 
For smaller $\alpha$ values the comparison is more problematic: in particular, while ED and PEPS confirm 
the existence
of an ordered spiral region for $\alpha$ around 1, the magnitude of the order parameter appears to be largely
overestimated by MSW theory, which is not surprising considering the partial account of quantum fluctuations by 
the MSW approach. In particular, MSW theory produces the counterintuitive prediction that the frustrated
spiral phase for $0.18 < \alpha < 1.35$ has an order parameter which can be larger (around the isotropic
$\alpha=1$ point) than that of the unfrustrated case of the square lattice (recovered for $\alpha\to\infty$).
We observe that  MSW theory is only moderately improving upon linear spin wave theory around 
the $\alpha=1$ point for what concerns the quantum fluctuations of the order parameter - in particular, 
its prediction for $M_0$ essentially coincides with that of LSW for $\alpha=1$. 
 
In summary, from the MSW order parameter $M_0$ we can derive a loss of LRO at $\alpha\lesssim 0.18$, and the spin stiffness suggests a strong weakening of inter-chain correlations already at $\alpha\lesssim 0.35$. The spin-stiffness also decreases strongly upon approaching the parameter region $1.35 \lesssim \alpha \lesssim 1.66$.
Together with the appearance of imaginary spin-wave frequencies for $\alpha\lesssim 0.18$ and $1.35 \lesssim \alpha \lesssim 1.66$ this strongly indicates the appearance of disordered phases in these regions. 
Due to its semiclassical character the MSW Ansatz is not adapted to properly describe these phases, and we must resort to 
methods such as PEPS which take quantum fluctuations into account more completely.
However, in the rest of parameter space magnetic LRO order seems to survive quantum fluctuations.

In the next section we investigate in detail the nature of the ordered phases.

\subsection{\label{cha:chirtriang}Ordering vector, spin--spin and chiral correlations}

In this section we introduce several observables which reveal the type of order appearing in the system, 
and which will be used in the discussions of sections~\ref{cha:transition2Dspiral} and~\ref{cha:msw_triangNN_1DNeel}.

The ordering vector is a direct outcome of MSW theory, and can be extracted from the ED data by determining the position of the peak of the static structure factor, Eq.~\eqref{Sk}. Figure~\ref{fig:Q_XY_triang} displays the $x$-component of the ordering vector $\vect{Q}$ (with $Q_y=0$).
Three limiting values are known. For $\alpha=0$ intra-chain antiferromagnetic (N\'{e}el) order is described by $\vect{Q}=\pi\hat x$. For $\alpha\to\infty$ square-lattice N\'{e}el order is described by $\vect{Q}=2\pi\hat x$. In the isotropic lattice ($\alpha=1$) the threefold symmetry forces the ordering vector to $\vect{Q}=\frac{4 \pi}{3}\hat x$. (The ED and PEPS results deviate at $\alpha=1$ because the required threefold symmetry is broken by the shape of the 
simulation clusters, Fig.~\ref{fig:systemsexactdiag}.) The importance of optimizing the ordering vector is apparent in Fig.~\ref{fig:Q_XY_triang} when comparing the MSW results to the classical (and LSW) curve.

The spin--spin correlations of nearest neighbors shed further light on the order properties.
We analyze them through the two-site total spin,
\begin{equation}
\label{totalspinTij}
K_{ij}\equiv\frac 1 2 \braket{\left(\vect{S}_i + \vect{S}_{j}\right)^2}=\braket{\vect{S}_i \cdot \vect{S}_{j}}+\frac 3 4\,.
\end{equation}
In Fig.~\ref{fig:SS_XY_triang} we plot it for nearest neighbors. This quantity vanishes if the spins are in a singlet, which is equivalent to perfect anticorrelation, takes the value~$\frac 3 4$ if they are uncorrelated, and the value~$1$ if the spins form a triplet, which means perfect correlation.
For PEPS and ED we report the values of $K_{ij}$ averaged over the central spins, where boundary effects are minimal.

Spiral phases carry not only a magnetic order parameter, but also a chiral order parameter. In particular, a vector chirality \cite{Kawamura2002} can be defined on an upwards pointing triangle with counter-clockwise labeled corners $\left(i,j,k\right)$ as
$
\kappa_\Delta=\frac 2 {3\sqrt{3}}\left[\vect{S}_i\times\vect{S}_j+\vect{S}_j\times\vect{S}_{k}+\vect{S}_{k}\times\vect{S}_i\right]_z,
$
and on a downwards pointing triangle with counter-clockwise labeled corners $\left(i,l,j\right)$ as
$\kappa_\nabla=\frac 2 {3\sqrt{3}}\left[\vect{S}_i\times\vect{S}_l+\vect{S}_l\times\vect{S}_{j}+\vect{S}_{j}\times\vect{S}_i\right]_z
$. 
Long-range chirality correlations are defined as \cite{Richter1991}
\begin{equation}
    \label{psi-}
    \psi_-=\braket{\left(\kappa_\Delta-\kappa_\nabla\right)\left(\kappa_{\Delta '}-\kappa_{\nabla '}\right)}\,,
\end{equation}
where the triangle pairs $\left(\Delta,\nabla\right)$ and $\left(\Delta ',\nabla '\right)$ share a $\tau_1$ edge.
In Fig.~\ref{fig:CC_XY_triang} we plot the average chirality correlation of the central plaquette with all other plaquettes, normalized to the theoretical maximum $4/9$. The MSW data have been obtained by expanding the chiral correlation up to the fourth order in the boson operators, 
which is consistent with the truncation of the bosonic Hamiltonian Eq.~\eqref{H4} to the same order. Going to higher orders does not change the outcome in the regions where $M_0$ is large, but can yield different results where $M_0$ is small. In particular, the unphysical negative values attained
by $\psi_{-}$ for small $\alpha$ are an artifact of this truncation.

\begin{figure}
        \begin{center}
        \null~~~~~~~~~\includegraphics[width=0.8\textwidth]{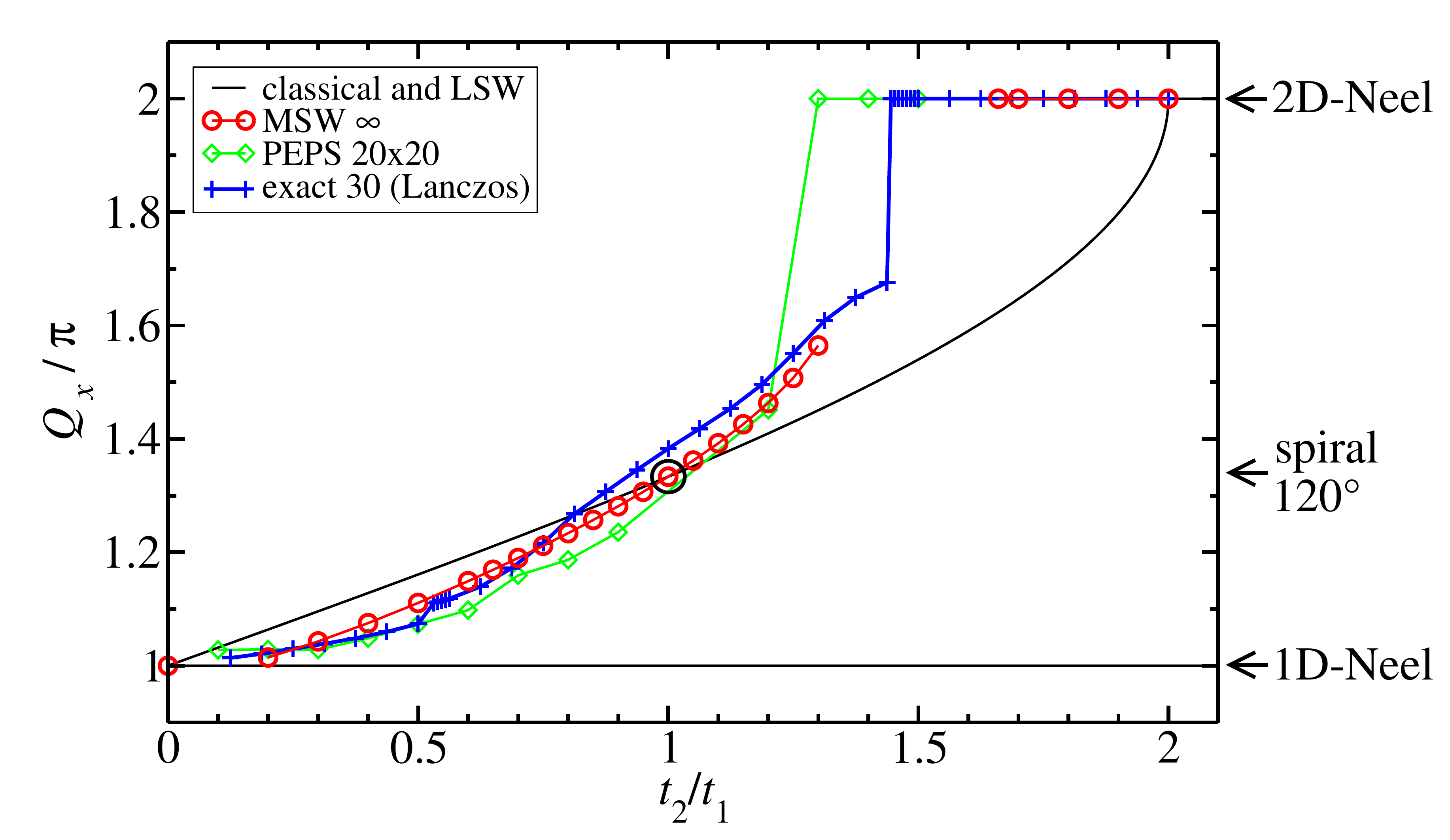}
        \caption{\label{fig:Q_XY_triang}
        Comparison of the $x$-component of the ordering vector, $Q_x$, using ED (blue), PEPS (light green) and the MSW Ansatz (red). Also shown are the classical values (black). The numbers in the labels of the curves are the respective system sizes. The black circle marks the isotropic spiral ordering vector of $Q_x=120^\circ$ which occurs classically and within MSW theory at $\alpha=1$.
        }
				\end{center}
\end{figure}
\begin{figure}
        \begin{center}
        \null~~~~~~~~~\includegraphics[width=0.8\textwidth]{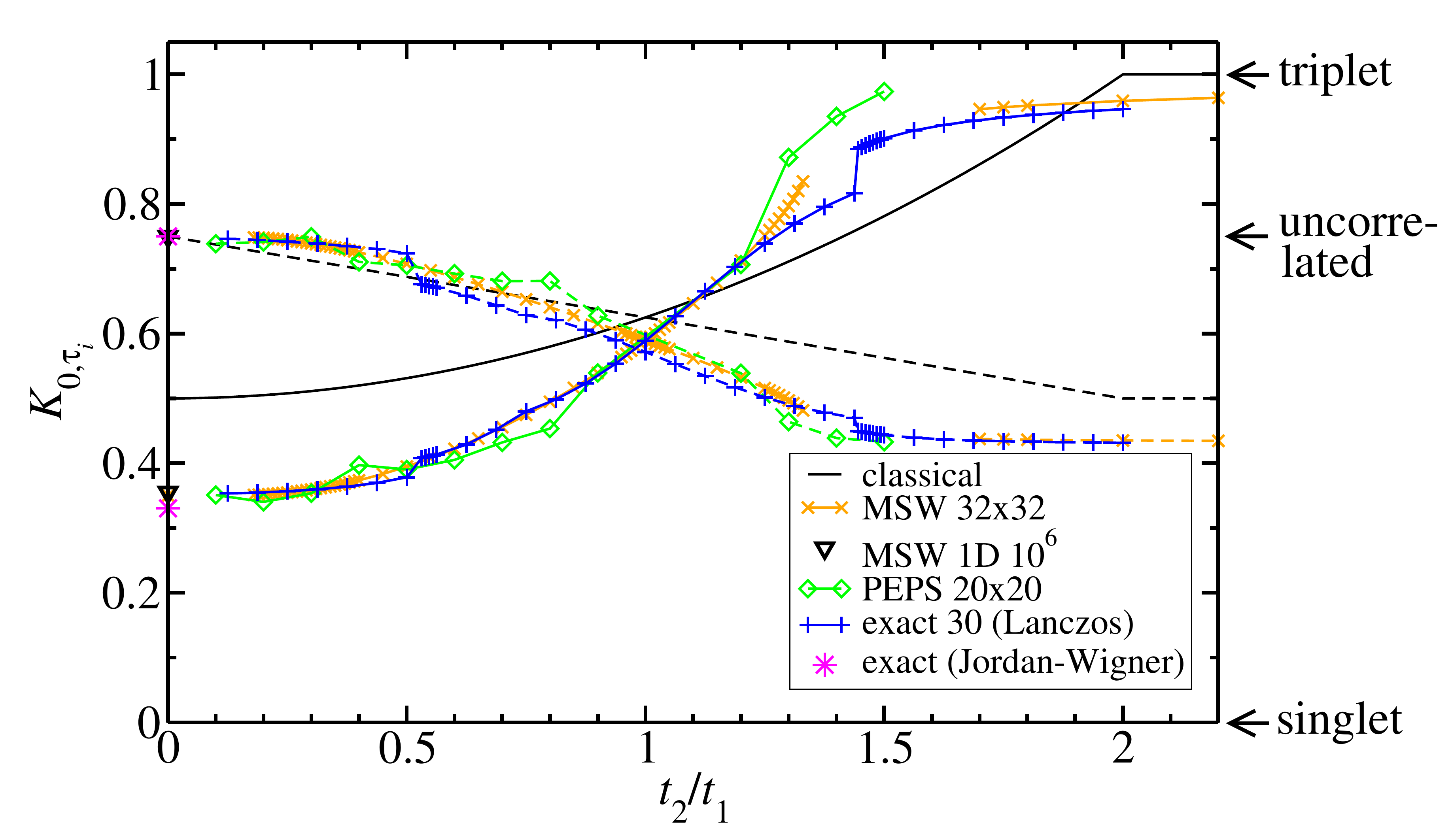}
        \caption{        \label{fig:SS_XY_triang}
        Nearest-neighbor correlation $K_{0,\vs{\tau}_i}$, where $\vs{\tau}_i=\vs{\tau}_1$ (solid lines) and $\vs{\tau}_i=\vs{\tau}_2$ (dashed lines), respectively, comparing ED (blue), PEPS (green), and MSW theory (orange). The black triangles are the MSW data for a one-dimensional chain of length $N=10^6$ and the stars in magenta are the exact results for a linear chain in the thermodynamic limit derived by use of the Jordan--Wigner transform.
        }
        \end{center}
\end{figure}
\begin{figure}
        \centering
    		\includegraphics[width=0.7\textwidth]{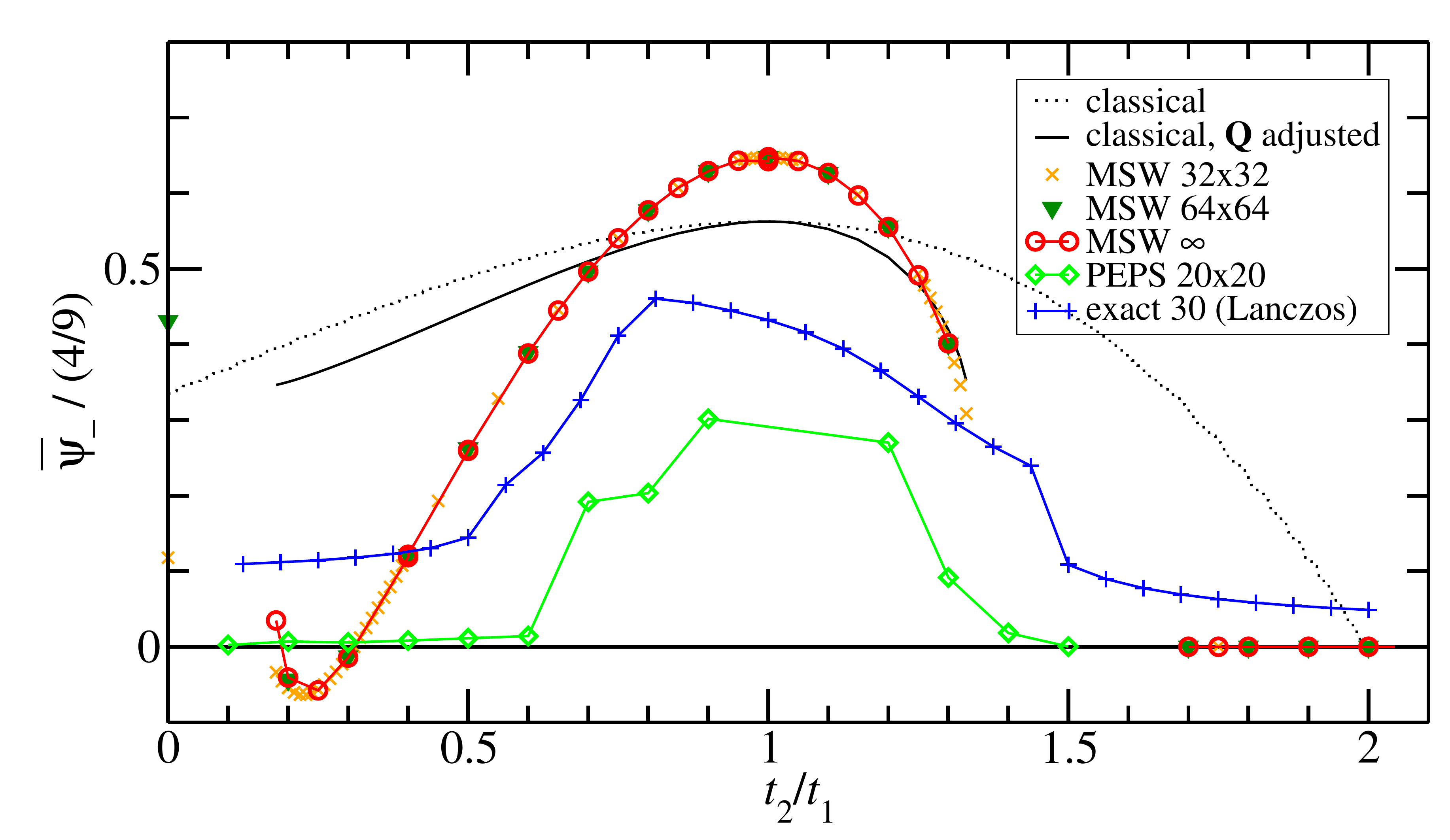}
        \caption{\label{fig:CC_XY_triang}
        Averaged chiral correlation normalized to the theoretical maximum $4/9$ for ED (blue), PEPS (light green), and MSW theory (orange, dark green, and red). The black dotted line is the classical result and the black solid line is the classical chiral correlation that is obtained if for a given $\alpha$ the $\vect{Q}$ of the MSW calculation rather than $\vect{Q}^{\mathrm{cl}}$ is used.
        }
\end{figure}

\subsection[Transition from 2D-N\'{e}el order to spiral order]{Transition from 2D-N\'{e}el order to spiral order\label{cha:transition2Dspiral}}
An inspection of Figs.~\ref{fig:Q_XY_triang},~\ref{fig:SS_XY_triang}, and~\ref{fig:CC_XY_triang} shows that the MSW formalism indeed reproduces the main 
features of the phase diagram of Fig.~\ref{fig:phasediagtriang}~(b) quite accurately.

First of all, coming from the large-$\alpha$ limit, we observe that all methods (MSW theory, ED and PEPS)
show a jump in the wavevector associated with the dominant correlations in the system,
from ${\bm Q}=2\pi \hat{x}$, characteristic of the N\'eel phase, to a continuously varying $\bm Q$,
characteristic of a spiral phase. 
PEPS indicates a jump from dominant N\'{e}el correlations to dominant
spiral correlations at $\alpha\approx 1.4$.
ED for 30 spins, shows a first order phase transition (with a sharp level crossing)
between N\'eel and spiral correlations at the slightly larger value of $\alpha\approx 1.44$. 
In the case of MSW theory, the jump in the ordering vector is realized when 
going across the breakdown region, namely when passing from $\alpha\approx1.66$ 
(which is the lower bound to the N\'eel phase within MSW theory) to 
$\alpha\approx1.35$ (which represents the upper bound of the spiral phase). 
In particular, all three different approaches point to the fact that N\'{e}el order persists to much 
lower $\alpha$ than the classical value $\alpha=2$.

The persistence of N\'{e}el order over a larger parameter region 
than in the classical case is reminiscent of what is observed in other models.
Indeed, quantum fluctuations generally stabilize states where spins are ordered 
collinearly (see \emph{e.g.\ }Refs.~\cite{Krueger2000, Henley1989}), a property that is reproduced by 
the MSW Ansatz with ordering vector optimization. The mechanism behind it is 
strongly connected to order-by-disorder phenomena \cite{Henley1989}.

The abrupt transition from a phase with dominant N\'eel correlations to a phase with
dominant spiral correlations is confirmed by the spin--spin correlations as displayed in Fig.~\ref{fig:SS_XY_triang}.
Anti-correlation along the strong $\vs{\tau}_2$-bonds and correlation along the weak $\vs{\tau}_1$-bonds are characteristic of a 2D-N\'{e}el ordered state; these correlations decrease rapidly for $\alpha < \alpha^{\mathrm{crit}}$ outside of the N\'{e}el-ordered phase. 
The change of the type of order is further supported by the overlap $\left|\braket{\psi_{\alpha}|\psi_{\infty}}\right|$ of the new ground state with the 2D-N\'{e}el ordered state of $\alpha=\infty$, which we plot in Fig.~\ref{fig:overlap} for the ED: above the phase transition it still attains a finite and quite large value, while it vanishes identically below the phase transition.
Finally, the onset of strong chiral correlations shows that the new phase is indeed a spirally 
correlated one (Fig.~\ref{fig:CC_XY_triang}). 

The breakdown region of MSW theory, $1.35 \lesssim \alpha \lesssim 1.66$, is strongly
suggestive of the loss of magnetic LRO, corresponding to a spin-liquid state. This region
of disordered behavior is only roughly consistent with that indicated by PEPS calculations
\cite{Schmied2008} for the appearance of a short-ranged spin-liquid phase, namely
$1.2 \lesssim \alpha \lesssim 1.4$. Nonetheless it is tempting to associate the breakdown of MSW
theory to this quantum-disordered phase. 
   
    \begin{figure}
        \centering
        \includegraphics[width=0.6\textwidth]{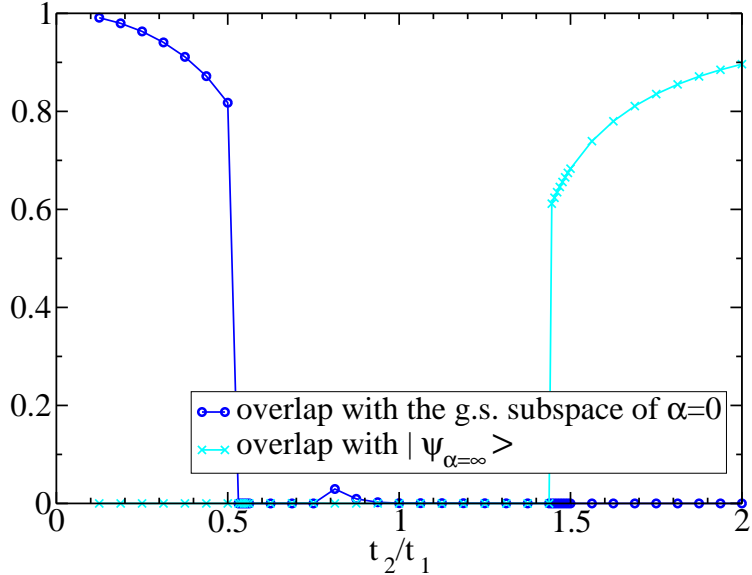}
        \caption{Overlap of the ground state at $\alpha$ with the 2D-N\'{e}el ground state of $\alpha=\infty$, $\ket{\Psi_{\alpha=\infty}}$, and with the six-dimensional subspace that corresponds to the ground state of $\alpha=0$, respectively (ED, 30 spins).
        }
        \label{fig:overlap}
    \end{figure}

\subsection{Persistence of 1D quasi-LRO up to finite inter-chain couplings\label{cha:msw_triangNN_1DNeel}}
In MSW theory, at $\alpha\approx0.18$ the order parameter $M_0$ breaks down. This is an indication of the transition to a phase without magnetic LRO such as
the one reproduced in the one-dimensional limit $\alpha\to 0$.
For $\alpha\gtrsim 0.18$ the weak $\alpha$-dependence of the ordering vector, the spin--spin correlations and the energy strongly indicate the persistence of a phase with 1D quasi-LRO properties to finite inter-chain couplings.
Furthermore, the spin stiffness component $\rho_{yy}$ practically vanishes for $\alpha \lesssim 0.35$ which suggests that, 
if we took quantum fluctuations into account in a more accurate way than in MSW theory, spiral (2D) order would be 
probably lost below $\alpha\lesssim 0.35$.

In the limit $\alpha\to 0$ we find that the intra-chain spin--spin correlation $K_{i,i+\vs{\tau}_1}\equiv K_{\vs{\tau}_1}\simeq 0.355$ lies close to the exact result that can be obtained by use of the Jordan--Wigner transform,
$K_{\vs{\tau}_1}=-\frac{1}{\pi}-\frac{1}{\pi^2}+\frac 3 4\simeq 0.330$.
The inter-chain spin--spin correlations on the other hand vanish, which is equivalent to $K_{i,i+\vs{\tau}_2}\equiv K_{\vs{\tau}_2}=\frac 3 4$.

We find that the ordering vector of the MSW theory compares very well to the one computed by ED in the entire range of $\alpha$ (Fig.~\ref{fig:Q_XY_triang}).
Especially the very weak dependence of $\vect{Q}$ on $\alpha$ near the 1D-limit is found in both approaches, contrary to classical and LSW theories which exhibit linear dependences on $\alpha$.
This means again that quantum fluctuations stabilize collinear order within the chains.
This is confirmed by ED: the overlap $\sqrt{\sum_{i=1}^6\left|\braket{\psi_{\alpha}|\psi_{0}^i}\right|^2}$ of the ground state with the subspace spanned by the six-fold degenerate \footnote{This degeneracy is due to the particular geometry of the 30-spin system, compare Fig.~\ref{fig:systemsexactdiag}. At $\alpha=0$ the even chains have to be in a singlet state while the four odd chains may each be in a state with total spin $\pm 1/2$, which yields a degeneracy of $2^4$. Restriction to the (physical) states with zero total magnetization $M_z$ results in a reduction of this degeneracy to a six-fold one.}
ground-states of $\alpha=0$ 
remains very large (almost 80 percent) up to $\alpha\approx 0.5$~(Fig.~\ref{fig:overlap}).
Moreover, the small chiral correlations in both PEPS and ED show that for small $\alpha$ 
there is no spiral long-range order.

\subsection{Momentum distribution of the hardcore bosons \label{cha:nk}}

 After characterizing the zero-temperature phase diagram of the spin model in the previous 
 sections, we now wish to make contact with the cold-atom implementation of such a model
 via hardcore bosons. The most common observable in cold-atom experiments is the 
 momentum distribution \cite{Bloch2008}, which exactly corresponds to the static structure factor
 for $S=1/2$ spins~\eqref{Sk}
 \begin{equation}
   \label{nk}
 n_b\left(\vect{k}\right)= \frac{1}{N} \sum_{i,j}e^{-i \vect{k} \cdot(\vect{r}_{i}-\vect{r}_j)}  \braket{b_i^{\dagger} b_j} = S\left(\vect{k}\right)
\end{equation}
via the spin-to-hardcore-boson mapping described in Sec.~\ref{cha:spinboson}. 
Figure~\ref{fig:nk} shows the MSW prediction for the momentum distribution at various 
$\alpha$ values, spanning all the condensation regimes of the bosons at zero
temperature. At $\alpha=0$ (not shown) the system displays quasi-condensation at finite
momenta along the uncoupled chains, resulting in vertical ridges at 
$Q_x = \pm \pi$ in the momentum distribution. These ridges  
corrugate as the interchain coupling increases, and true condensation 
peaks emerge in reciprocal space, corresponding
to a condensate state which supports a crystalline vorticity pattern. For $\alpha=1$ these peaks are located at the six corners of the first Brillouin zone. 
For $\alpha<1$ the peaks are elongated in the $y$ direction, while 
for $\alpha>1$ they are elongated in the $x$ direction, witnessing the 
spatial anisotropy of the lattice. This situation persists up to the breakdown
of MSW theory at $\alpha=1.35$; after recovery of the theory 
at $\alpha=1.66$, the momentum distribution shows condensation at the four
corners of the Brillouin zone of a (deformed) square lattice, 
defined by the dominant diagonal bonds of the spatially anisotropic
triangular lattice.  

The peak height (normalized to the number of sites) is given by the square of the order parameter $M_0$.

 \begin{figure}
                \includegraphics[width=0.875\textwidth]{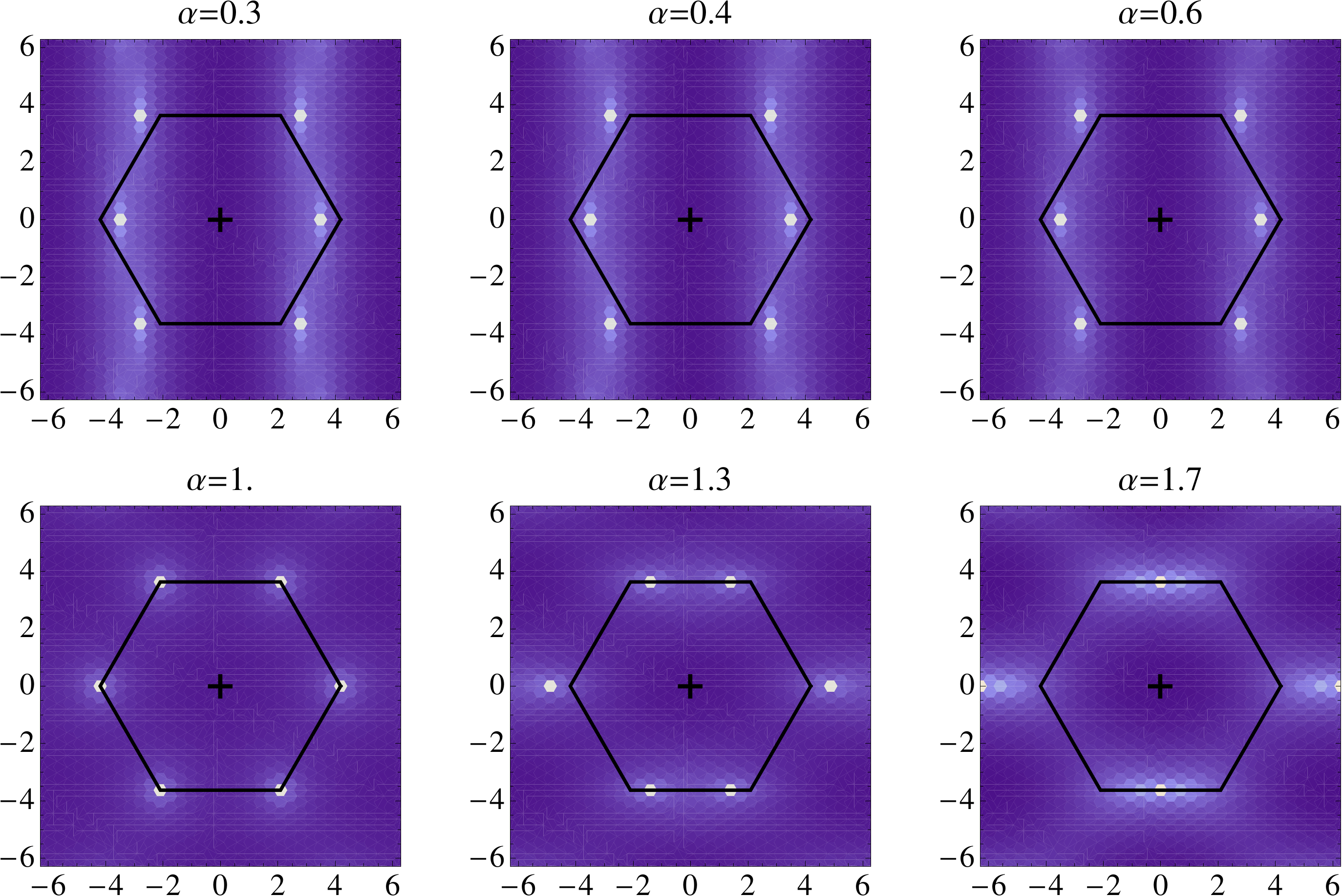}\qquad\includegraphics[height=4.5cm]{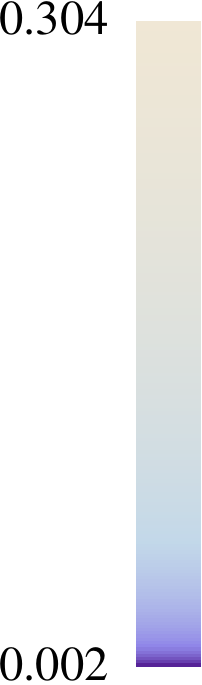}
        \caption{Momentum distribution of a half-filled gas of frustrated hardcore bosons on the spatially anisotropic triangular
        lattice [corresponding to the spin static structure factor $S(\vect{k})$], for various values of the spatial anisotropy $\alpha$. 
        The data (on a logarithmic color scale, scaled to the number of sites) result from a MSW calculation on an $18\times 18$ lattice. The black hexagon marks the first Brillouin zone and the black cross its origin.}
        \label{fig:nk}
    \end{figure}

\subsection{Discussion}

Here we summarize the main features of the zero-temperature phase diagram obtained via MSW theory with $\vect{Q}$-vector optimization.
The region where the system behaves like a set of essentially decoupled chains with 1D quasi-order is extended to considerable inter-chain interactions. The order parameter indicates that the point where inter-chain correlations set in occurs at $\alpha\approx 0.18$; the spin stiffness suggests that an effective decoupling of the chains may even persist up to $\alpha\approx 0.35$.
At larger $\alpha$ the system crosses over to a spirally ordered phase that persists up to
 $\alpha\approx1.35$, where MSW breaks down, suggesting a quantum disordered
 ground state. At $\alpha\approx 1.66$ MSW theory finds again a self-consistent
 solution, this time corresponding to a 2D-N\'{e}el state.

These results are mostly consistent with the PEPS phase diagram of Fig.~\ref{fig:phasediagtriang}~(b). Especially the persistence of  
1D behavior to surprisingly large values of $\alpha$, the fact that the long-range ordered spiral phase survives quantum fluctuations, and the extension of 2D-N\'{e}el LRO down to much smaller values of $\alpha$ than in the classical equivalent are reproduced.
However, there are some deviations, 
which are generally to be expected from a spin-wave approach. The range of 
the ordered phases appears to be somewhat overestimated by 
MSW theory. 
Furthermore the gapped spin-liquid phases are not faithfully described: while the breakdown of MSW theory
for $1.35 \lesssim \alpha \lesssim 1.66$ suggests a disordered ground state, 
the gapped spin liquid in the region $0.4\lesssim\alpha\lesssim0.6$, suggested by the PEPS data \cite{Schmied2008}, is not observed in the MSW results. 
Still the proposed phase diagram is significantly improved with respect to the LSW phase diagram, 
which follows the classical phase diagram too closely.

Two improvements have proven to be crucial: First, the minimization of the free energy with respect to $\vect{Q}$ in the self-consistent equations has enabled us to describe the considerable shift of the ordering vector to a surprisingly satisfactory level.
Second, the investigation of the Gaussian spin stiffness serves as a measure of the actual stiffness of the long-range ordered phase. As such it allows us to detect regions where spin-liquid behavior may appear in the true quantum ground state.

\end{section}


\begin{section}{\label{cha:msw_triangFiniteT}
Finite temperature phase diagram of the anisotropic triangular lattice
}

We now investigate how the phase diagram of the antiferromagnetic nearest-neighbor (NN) XY Hamiltonian on the anisotropic triangular lattice extends to finite temperatures, making use of the modified spin-wave (MSW) theory with ordering-vector optimization.
All calculations in this section are carried out in the thermodynamic limit. 

At finite temperatures, continuous symmetries cannot be spontaneously broken in two dimensions 
\cite{Mermin1966,Hohenberg1967}. In the XY model, instead of long-range order (LRO), one finds quasi-LRO at finite but low temperature. 
At the Berezinskii--Kosterlitz--Thouless (BKT) temperature $T_{\mathrm{BKT}}$ the system undergoes a topological phase transition from quasi-LRO to an exponential decay of correlations, 
involving the unbinding of vortex--antivortex pairs \cite{Berezinskii1971,Berezinskii1972,Kosterlitz1973}.  
The existence of a BKT transition in the XY model must be seen in contrast to the Heisenberg model, where Kosterlitz and Thouless showed that vortex excitations are not topologically stable \cite{Kosterlitz1973}, a fact which precludes the possibility of a BKT transition.

The possibility of observing the BKT transition is a particular advantage of MSW theory. The BKT phase with algebraic order 
is generally predicted by linear spin-wave (LSW) theory to remain stable at arbitrary temperatures. The non-linearities contained in MSW theory allow the disruption of quasi-LRO and the transition to the short-range-ordered (SRO) phase. However, vortex--antivortex excitations are not explicitly present in the theory, and therefore in principle
$T_{\mathrm{BKT}}$ cannot be accurately estimated.

\subsection{Spin--spin correlations \label{cha:finiteT_observables}}

An important observable for the analysis of a temperature-dependent phase diagram is the two-point correlation function
\begin{equation}
\label{corr}
C_{ij}\equiv\braket{S_i^x S_j^x+S_i^y S_j^y}/\cos\left(\vect{Q}\cdot\vect{r_{ij}}\right)=\frac 1 2 \left( F_{ij}^2+G_{ij}^2\right)\,.
\end{equation}
In our analysis we focus on 
$C_{m\, \vs{\tau}_{1}}$ and $C_{m\, \vs{\tau}_{2}}$
where $m$ is a positive integer, and $\vs{\tau}_1=\left(1,0\right)$ and $\vs{\tau}_2=\left(1/2,\sqrt{3}/2\right)$ are the lattice vectors. The behavior of $C_{m \, \vs{\tau}_{1}}$ captures the intra-chain correlations, while that of $C_{m \, \vs{\tau}_{2}}$ describes inter-chain correlations.

In order to locate the BKT transition we calculate the residual sum of squares $R=\sum_m \left[ C_{m \, \vs{\tau}_{1,2}} - f(m \, \vs{\tau}_{1,2}) \right]^2$ for two trial functions, an exponential $f(r)=A \ue^{-r/\xi}$, where $\xi$ is the correlation length, and an algebraic fit $f(r)=A/r^\eta$.
We fit these functions to the correlations of the central spin with sites which are $m=3\ldots 15$ lattice spacings apart. The lower limit to the fit region is necessary because the trial functions are only valid for the long-distance part of the correlations, while the upper limit is chosen by us because computation time for the correlations increases considerably with the distance between the spins. 
We identify a BKT transition with a point where the residual sum of squares $R$ of the exponential fit becomes equal to that of the algebraic fit. This has to be understood as only a rough estimate of the transition temperature.
When giving explicit values of transition temperatures we take the average of the values obtained from fits to $C_{m \, \vs{\tau}_{1}}$ and $C_{m \, \vs{\tau}_{2}}$.

Figure~\ref{fig:Cmt12} shows representative log--log plots of the correlation function $C_{m \, \vs{\tau}_{1}}$ at $\alpha\equiv t_2/t_1=0.7$ and $\alpha=100$ (where the ground states show spiral and 2D-N\'{e}el order, respectively) for several temperatures. In these plots algebraically decaying correlations correspond to straight lines.
    \begin{figure*}
        \centering
        \includegraphics[width=0.45\textwidth]{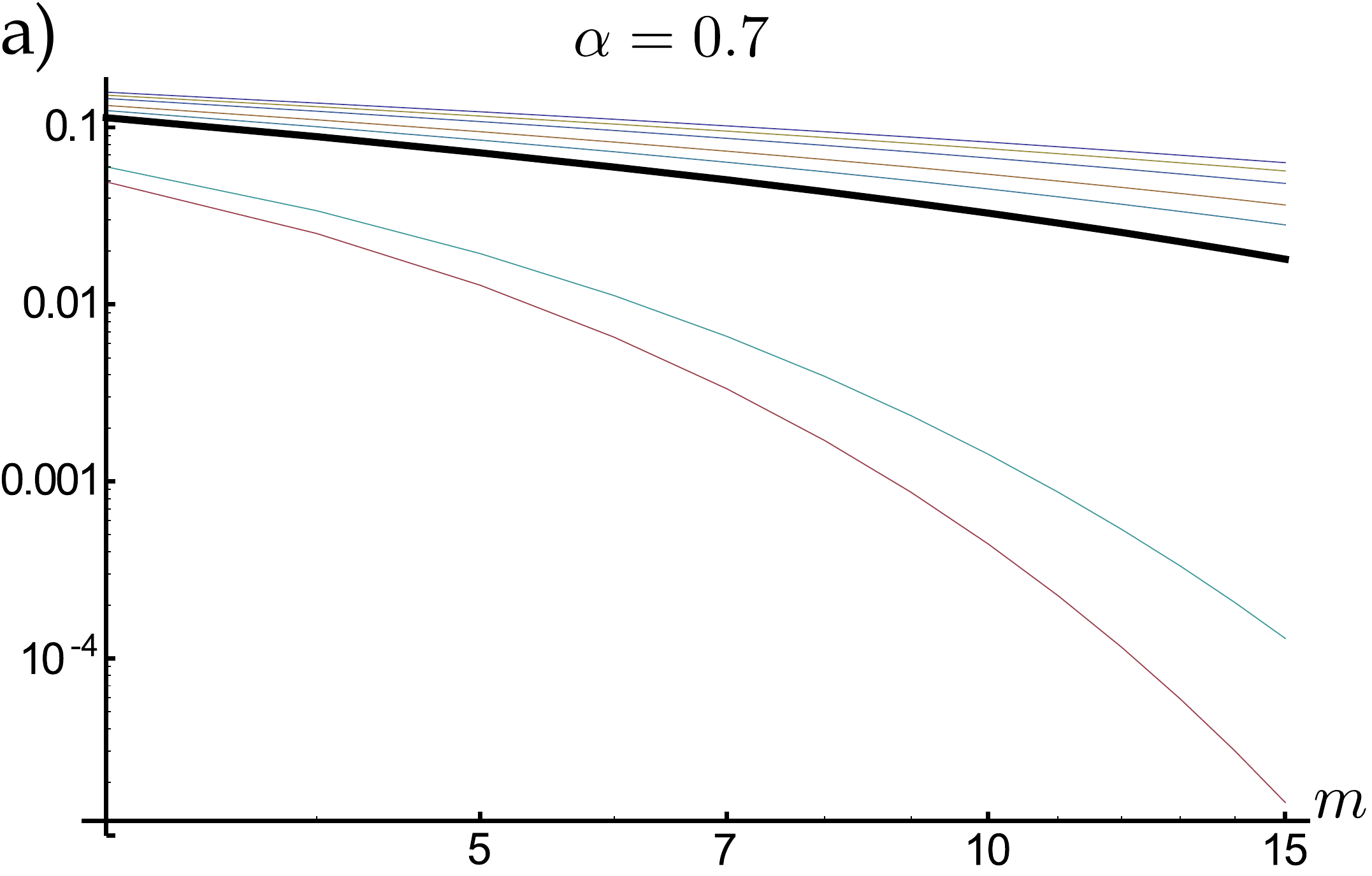}\quad  \includegraphics[width=0.45\textwidth]{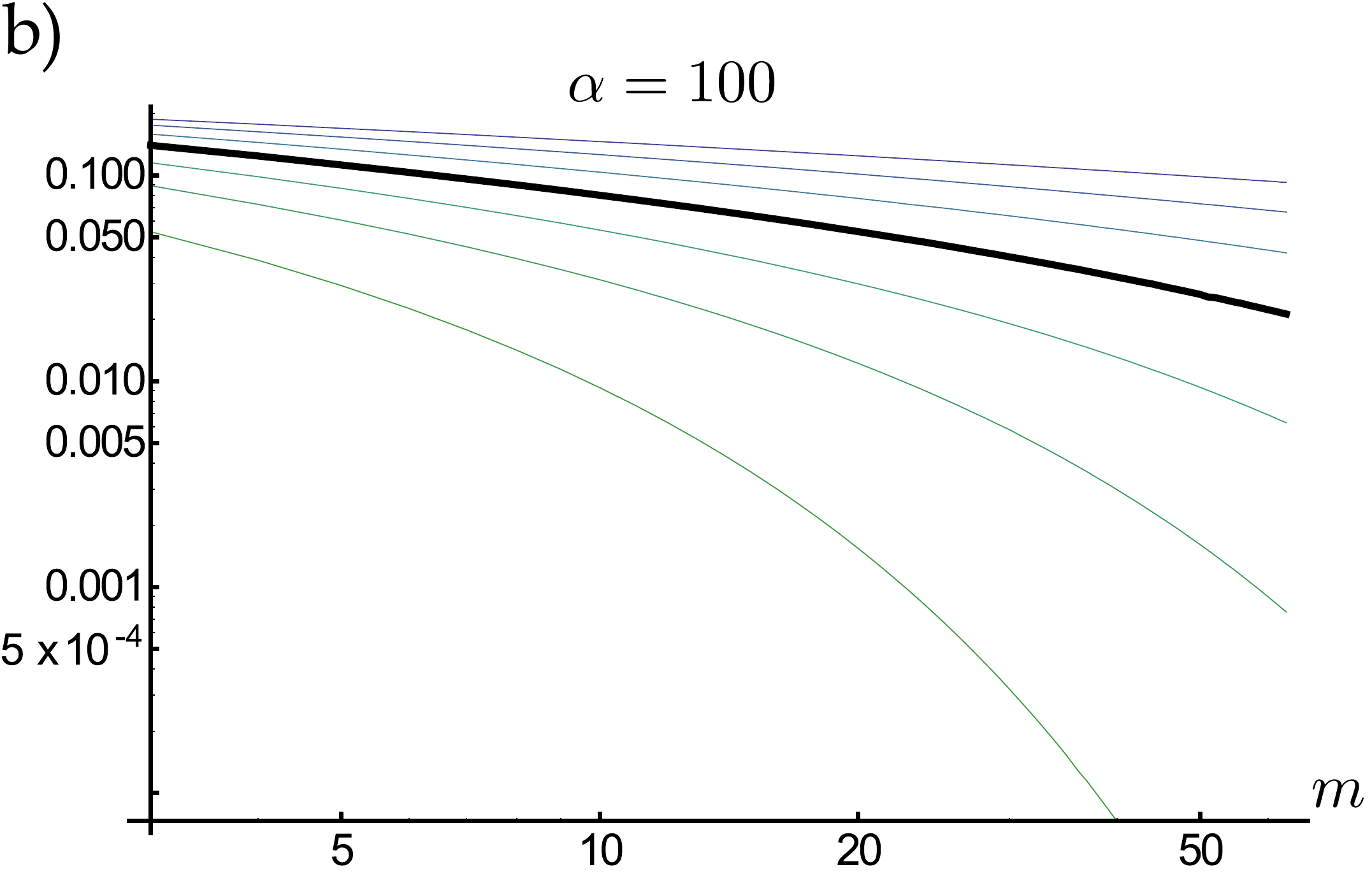}\\      
        \caption{Correlation function $C_{m \, \vs{\tau}_{1}}$ along the chains in the thermodynamic limit, (a) for $\alpha=0.7$, and (b) for $\alpha=100$ for several temperatures. 
        The normalized temperatures $T/(t_1+2 t_2)$ considered (for lines from top to bottom) are (a) $T/(t_1+2 t_2)=0.064\ldots 0.078$ in steps of $0.002$, and (b) $T/(t_1+2 t_2)=0.068\ldots 0.164$ in steps of $0.016$. The line closest to $T_{\mathrm{BKT}}$ calculated from MSW data is shown bold. Note the difference in the maximal $m$ between~(a) and~(b). }
        \label{fig:Cmt12}
    \end{figure*}
For  $\alpha=0.7$ a clear transition from algebraic to exponential decay at the computed BKT temperature can be observed, a behavior which is found in the entire parameter range of the spiral phase.
On the contrary, at $\alpha=100$ we cannot find such a clear transition. Rather, the curves acquire a curvature in a fairly continuous way, which makes it difficult to pinpoint the transition.
In order to check how the suggested BKT line changes when taking correlations to more distant spins into account, we computed spin--spin correlations for $\alpha=100$ up to distances of 64 lattice spacings. This yields a transition temperature of $T_{\mathrm{BKT}}/(t_1+2 t_2)=0.134$, which is approximately $15\%$ lower than what is obtained if distances of only up to 15 lattice spacings are considered. 
In light of the approximate nature of MSW theory, we find that this level of precision is satisfactory.

Another useful observable is represented by the gap $\Delta=\Delta_{\vect{k}=0}$ of the spin-wave dispersion, 
which is intrinsically connected to the two-point correlations. The gap is directly imposed by the chemical potential $\mu$ 
[see Eqs.~\eqref{AkBk} and~\eqref{disp}], and its magnitude determines the rapidity of the decay of correlations. 
A finite gap leads to exponentially decaying correlations while a vanishing one entails power-law correlations. 
Hence in principle the onset of a gap at finite temperatures corresponds to the occurrence of a BKT transition.

In reality, the thermal onset of a gap we observed within MSW is typically very gradual, 
and a clear identification of the transition point via the gap is generally problematic.
This observation can be understood on the basis of a well-known fact: the chemical
potential of the half-filled DM boson gas, which determines the existence of a gap, 
cannot vanish at finite temperature because of the absence of Bose-Einstein
condensation in two dimensions. As a consequence, we find a finite gap at any finite temperature, 
which means that the correlations decay exponentially at long distances. 
This suggests that, strictly speaking, MSW theory is not able to describe 
the BKT transition. However, for temperatures much lower than the BKT transition 
(estimated as explained above) the gap is very small, being below our numerical precision. 
For all practical purposes such a small gap entails a decay of correlations which 
is not distinguishible from an algebraical decay. Moreover in a selected region of the phase
diagram (corresponding to the spiral phase) the gap is seen to increase drastically
around the estimated BKT transition temperature, and correspondingly the correlation
function is seen to decay much more rapidly above that temperature. Hence we 
conclude that MSW theory still accounts for one of the most salient features
of the BKT transition, namely a discontinuous behavior of correlations
as the temperature is increased.

Finally we can extract from the correlations the temperature at which the MSW formalism breaks down.
It is characterized by the complete loss of all correlations, even to the nearest neighbor. 
This behavior, occurring at temperatures of the order of the coupling strength, is  clearly
an artifact of the method, since in real systems the complete loss of correlations occurs only at extremely large temperatures 
where spin--spin interactions become negligible.

\subsection{The phase diagram\label{cha:msw_triangFiniteTphd}}

In this section we present the finite-temperature phase diagram of the anisotropic triangular lattice model obtained via the MSW method with ordering vector optimization \footnote{Note that we use units in which the Boltzmann constant $k_\mathrm{B}$ equals unity.}.
We derive it from the observables introduced in the previous section. 
For reference we first present a summarizing sketch of the phase diagram in Fig.~\ref{fig:TLXYfiniteTPhD}, which introduces the phase labels we will refer to in the following discussion. Table~\ref{tab:PhasesFiniteT} lists the main properties of these phases.
\begin{figure}
				\centering
				\begin{turn}{270}
        \includegraphics[width=0.35\textwidth]{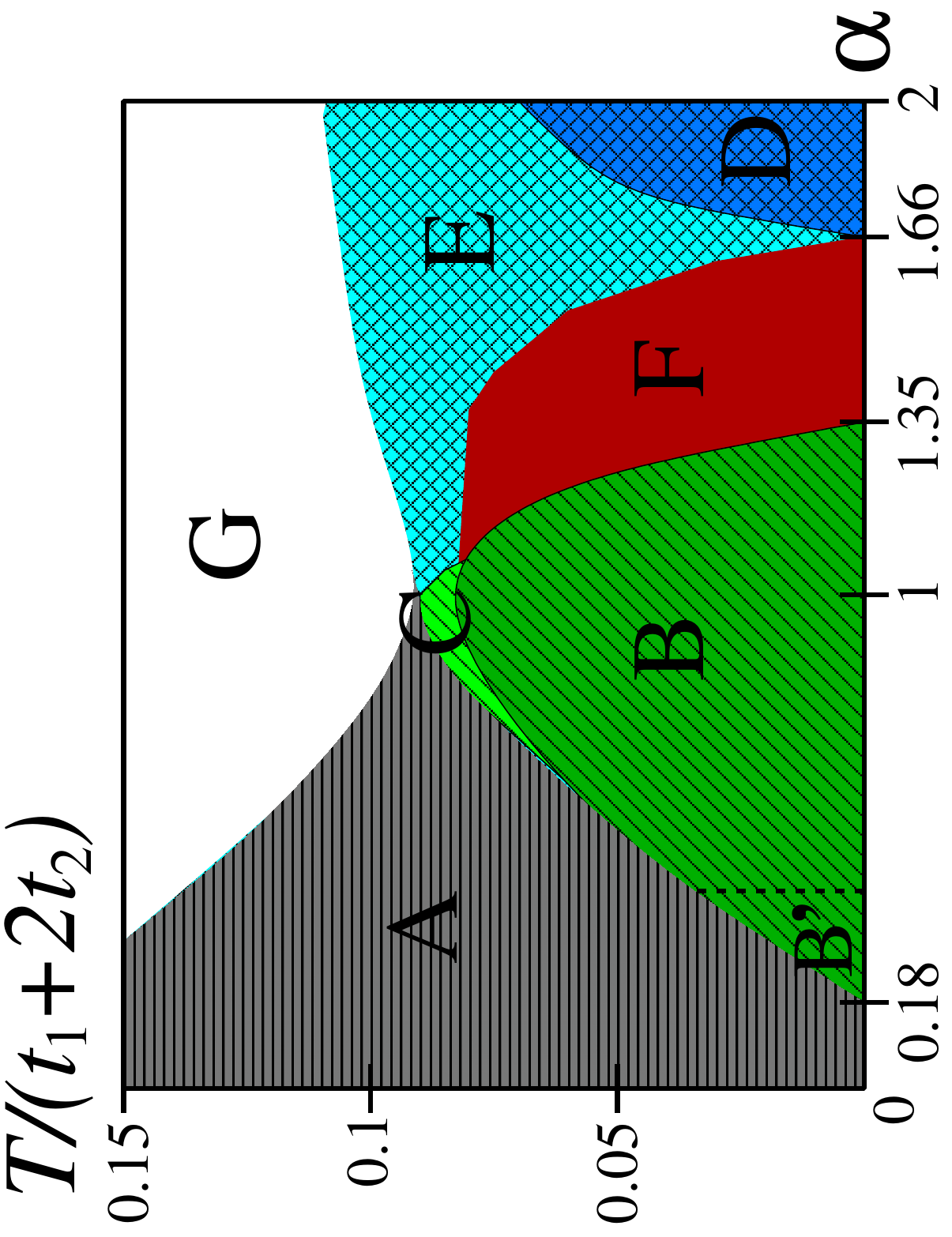}
        \end{turn}

        \caption{
        Schematical temperature-dependent phase diagram of the XY SATL. The different regions are listed along with their main characteristics in table~\ref{tab:PhasesFiniteT}. Horizontal lines mark 1D-Néel order; diagonal lines spiral order; and cross-hatches 2D-Néel order. \label{fig:TLXYfiniteTPhD}
        }
    \end{figure}

\begin{table*}
		\centering
		\begin{tabular}{|ll||l|l|}
  		\hline
  		Phase &															& $Q_x$ 					& decay of correlations \\
  		\hline
  		(A)		& 1D-like SRO			 						&	$\pi$						& intra-chain: exponential	\\
  					&															&									& inter-chain: uncorrelated	\\
  		(B)		& Spiral quasi-LRO						& $\pi<Q_x<2\pi$ 	& algebraic							\\
  		(C)		& Spiral SRO									& $\pi<Q_x<2\pi$ 	& exponential						\\
  		(D)		& 2D-N\'{e}el quasi-LRO				& $Q_x=2\pi$	 		& algebraic							\\
  		(E)		& 2D-N\'{e}el SRO							& $Q_x=2\pi$	 		& exponential						\\
  		\hline
  		(F)		& Unstable (imaginary modes)	& ---					 		& ---										\\
  		(G)		& Breakdown of theory					& ---					 		& no correlations				\\
  		\hline
  	\end{tabular}
  	\caption{\label{tab:PhasesFiniteT}Parameter regions found in the finite-temperature phase diagram of Fig.~\ref{fig:TLXYfiniteTPhD}. We distinguish mainly between phases with quasi-long-range order (quasi-LRO), \emph{i.e.\ }algebraic decay of correlations, and phases with short-range order (SRO), \emph{i.e.\ }exponential decay of correlations. Moreover, two regions are listed where the MSW formalism ceases to be applicable [(F)~and~(G)].}
\end{table*}

First, at small $\alpha$, there is a phase with properties similar to the algebraic 1D-N\'{e}el-like state found at $T=0$ but with exponential decay of intra-chain correlations (phase A). 
Further, we generally find that the phase diagram contains two quasi-LRO regions: 
a region at intermediate $\alpha$ corresponding to spiral quasi-LRO (phase B), and another region at large $\alpha$ which is characterized by N\'eel quasi-LRO (phase D). 
These phases undergo BKT transitions to similar phases with short-range order (SRO), phases C and E, respectively.
Moreover, between them lies a region where imaginary frequencies occur in the spin-wave dispersion relation, which can be interpreted as an indication for an extremely short-range-ordered phase (phase F).
This general structure of the phase diagram is supported by all the observables we investigate. 
At large $T$ the MSW method breaks down (see sec.~\ref{cha:finiteT_observables}) and therefore does not allow for any interpretation in that domain (region G).
         
It is also important to note that our calculations cease to converge properly for too low temperatures, when the chemical potential becomes smaller than the accuracy of our numerical integrations. 
Depending on the region of the phase diagram the lowest temperatures for which appropriate results could be derived vary from less than one-tenth of a percent to several percent of the coupling strengths. 
This pathology is not observed at $T=0$ (as calculated in section~\ref{cha:msw_triangNN}) because an exact vanishing of the chemical potential allows the special treatment of the zero-mode as captured in Eqs.~\eqref{FG} and~\eqref{constr2}. Except for some points, we typically calculated down to $T/(t_1+2t_2)=0.025$. 
Since the bond strengths are the only energy scales in the problem, it seems a reasonable assumption that our finite temperature results can be analytically continued down to $T=0^+$ without encountering discontinuities (except possibly at exactly $T=0$, where --- in contrast to any finite temperature --- Bose--Einstein condensation of the 
DM bosons becomes possible). Nevertheless, this issue should be kept in mind in the following analysis. 

The breakdown of the calculations for too low temperature can be clearly seen in Fig.~\ref{fig:phd_finiteT_MSW}, which displays the phase diagrams obtained from several observables. 

A natural starting point for a thorough analysis of the temperature dependent phase diagram is given by the respective ground state phases. Proceeding from small to large $\alpha$, we divide the analysis in sections corresponding to different ground state behavior.

    \begin{figure*}
        \centering
        \includegraphics[width=0.33\textwidth]{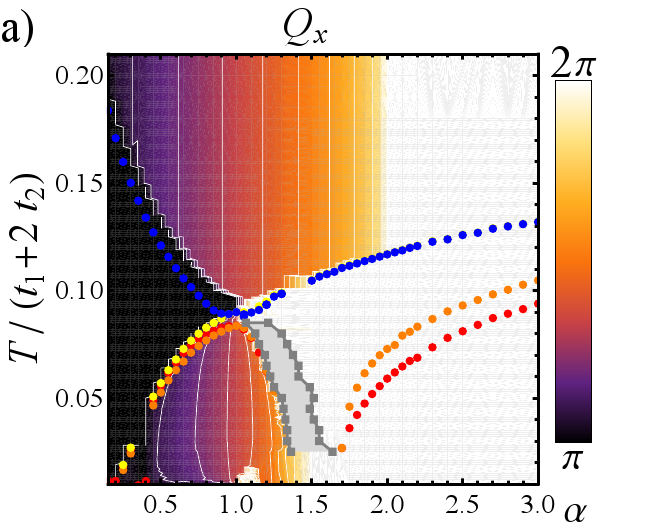}\hspace*{0.2cm}
 \includegraphics[width=0.33\textwidth]{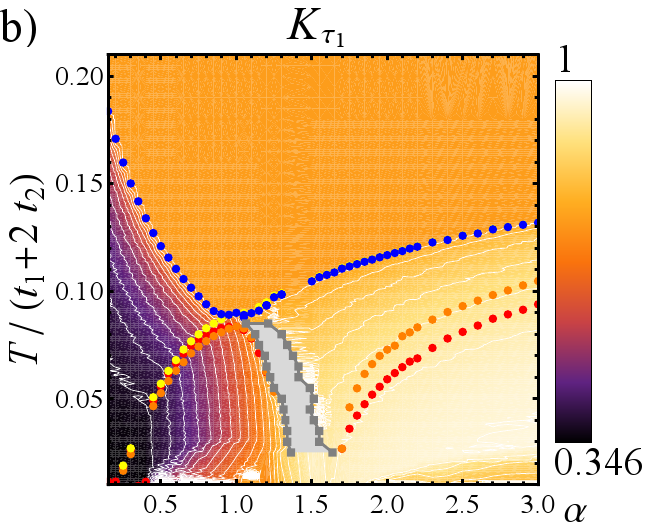}\hspace*{0.2cm} 
  \includegraphics[width=0.33\textwidth]{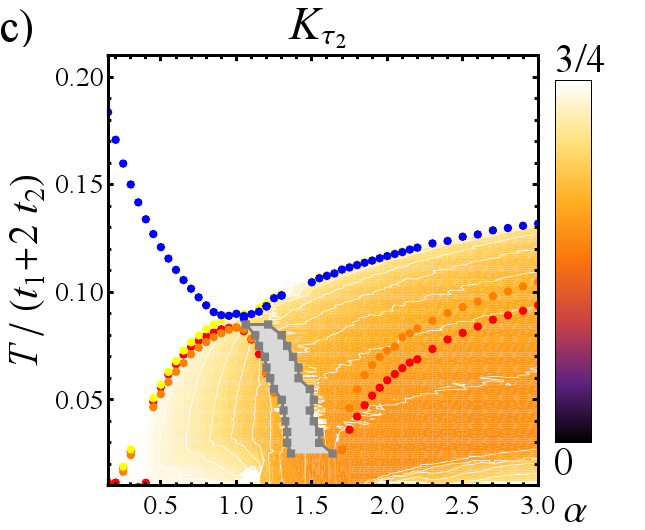}
        \vspace{0.3cm}

        \includegraphics[width=0.33\textwidth]{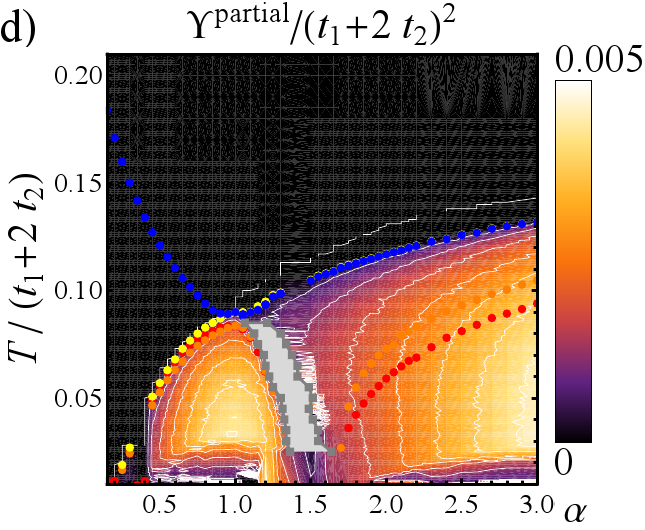}\hspace*{0.2cm}
  \includegraphics[width=0.33\textwidth]{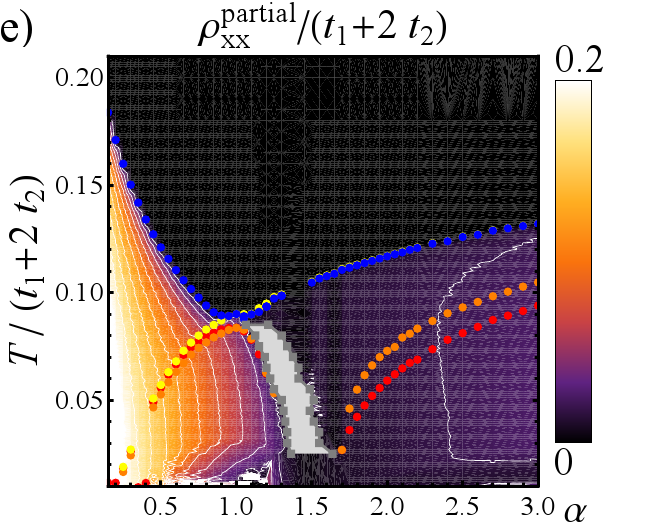}\hspace*{0.2cm}
  \includegraphics[width=0.33\textwidth]{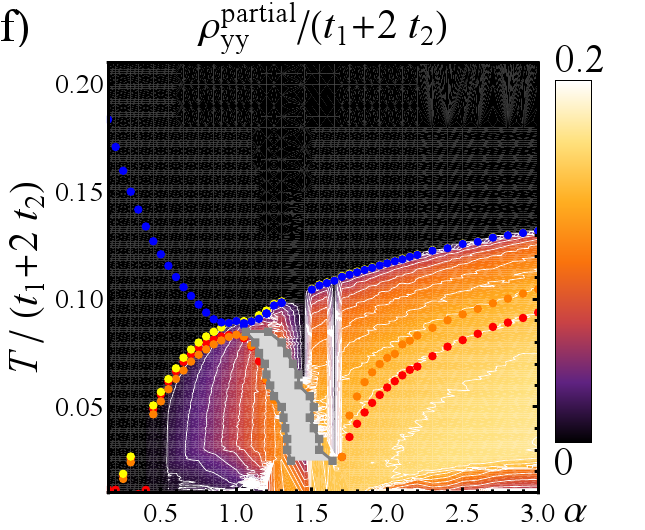}
  
  \vspace{0.5cm}
  \includegraphics[width=0.33\textwidth]{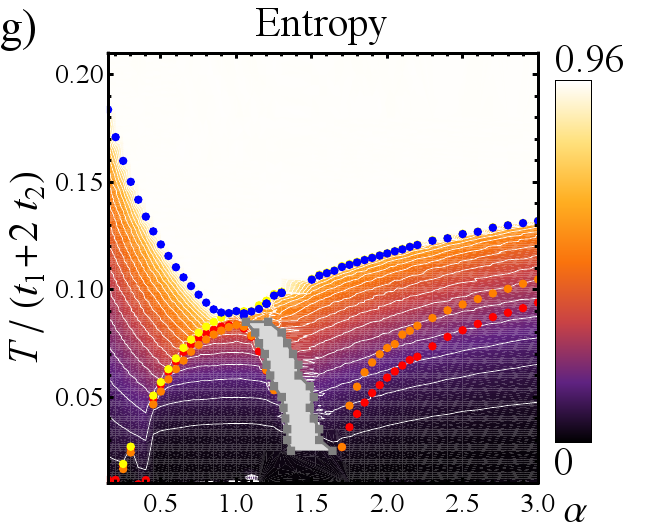}\hspace*{0.2cm}
 	\includegraphics[width=0.33\textwidth]{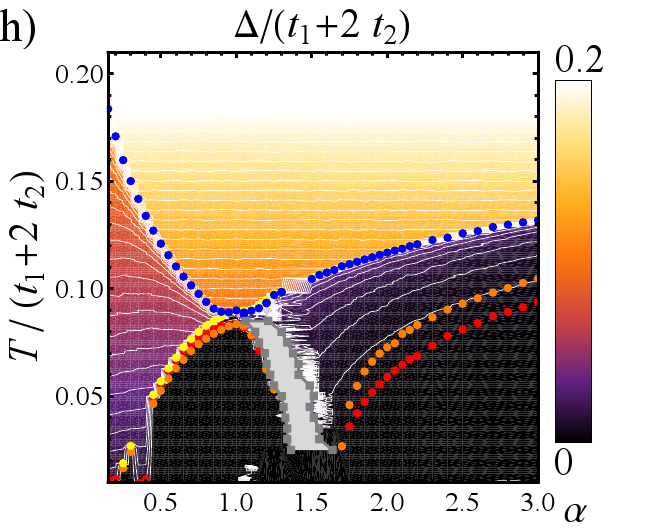}\hspace*{0.2cm}
 	\hspace*{0.33\textwidth}

        \caption{\label{fig:phd_finiteT_MSW}Linear color plots in dependence of $\alpha$ and $T/\left(t_1+2t_2\right)$ of (a) the $x$-component of the ordering vector $Q_x$, (b) the intra-chain correlation $K_{\vs{\tau}_1}$, (c) the inter-chain correlation $K_{\vs{\tau}_2}$, (d) the partial Gaussian spin stiffness $\Upsilon^{\mathrm{partial}}$, and~(e) and~(f) the partial spin stiffnesses $\rho_{xx}^{\mathrm{partial}}$ and $\rho_{yy}^{\mathrm{partial}}$, respectively. The mixed component of the spin stiffness $\rho_{xy}^{\mathrm{partial}}$ vanishes for symmetry reasons. 
        Figures~(g)~and~(h) show the entropy and the gap $\Delta$.
        The points mark the BKT transition for $\vs{\tau}_1=\left(1,0\right)$ (red), the BKT transition for $\vs{\tau}_2=\left(1/2,\sqrt{3}/2\right)$ (orange), the break-down temperature (blue), and the temperature where inter-chain correlations disappear (yellow), all computed through the two-point correlations $C_{m\, \vs{\tau}_{1,2}}$.
        In the grey region imaginary frequencies appear in the spin-wave dispersion.}
    \end{figure*}

\subsubsection{1D-like phase (phase A)}
The one-dimensional quasi-ordered ground state phase for which we found strong indications below $\alpha\approx 0.18$ becomes a short-range-ordered phase at finite temperature (phase A). It is characterized by vanishing correlations between neighboring chains already at zero temperature.
The finite gap leads to exponentially decaying intra-chain correlations for all $T$. This is consistent with the expected finite-temperature behavior above a ground state with quasi-LRO.

The assumption that this low-$\alpha$ phase really describes decoupled chains is reinforced by the component $\rho_{yy}^{\mathrm{partial}}$ of the spin stiffness, which vanishes in this region, and by the ordering vector that takes on the 1D value $\left(\pi,0\right)$, similar to the equivalent of the ground state phase diagram. Moreover, neighboring spins on different chains are uncorrelated whereas nearest-neighbors on the same chain are anti-correlated. 
It is remarkable that this phase is preferred over the quasi-ordered spiral phase with rising temperatures. In section \ref{cha:msw_triangNN_1DNeel} we have seen that quantum fluctuations stabilize 1D-N\'{e}el quasi-order. The same mechanism is at work here: collinear spin correlations are stabilized by fluctuations, in this case thermal ones.

Note that in the 1D-like phase A the inter- and intra-chain correlations behave completely differently. In the rest of the phase diagram they follow one and the same pattern, since in a truly two-dimensional structure the correlations in one direction typically cannot disappear without affecting the correlations in the other one.

\subsubsection{Spiral phases (phases B and C)}

At intermediate inter-chain couplings $0.18\lesssim\alpha\lesssim 1.35$ and low temperature we find a spiral phase with magnetic quasi-LRO (phase B). It can be seen as the finite-temperature continuation of the spirally ordered ground state phase. 
At larger temperatures a BKT transition to a phase with a spiral ordering vector but with an exponential decay of correlations occurs (phase C). 
Within our resolution of the phase diagram it seems that phase C disappears on the large-$\alpha$ side of the B-phase dome, 
and that phase B is delimited at large $\alpha$ by the F region where imaginary spin-wave frequencies appear.
Furthermore, on the low-$\alpha$ side phase C becomes extremely narrow and is almost immediately followed by a transition to phase A described in the previous section. The broadest extent in temperature of C is around $\alpha=1$.

At the isotropic point $\alpha=1$ the BKT transition from B to C is approximately located at $T_{\mathrm{BKT}}/(t_1+2 t_2)=0.0836$. 
Quantum effects lower the transition temperature considerably from the classical value $T_{\mathrm{BKT}}^{\mathrm{cl}}/(t_1+2 t_2)=0.165$ found by classical Monte Carlo simulations \cite{Lee1984}.
A pure-quantum self-consistent harmonic approximation, developed in Ref.~\cite{Capriotti1999b}, 
gives $T_{\mathrm{BKT}}/(t_1+2 t_2)=0.0625$. 
The fact that MSW theory gives a significantly higher estimate is not surprising given that Ref.~\cite{Capriotti1999b} takes vortex--antivortex excitations explicitly into account while MSW theory does not. 

We also find that in the same domain of large frustration ($\alpha$ close to 1) where spiral quasi-order is most stable, the breakdown of the theory occurs at a lower critical temperature.

We note also a strong drop of the transition temperatures around $\alpha\approx 0.4$. In Fig.~\ref{fig:TLXYfiniteTPhD} this is marked by a dashed line which separates phase B from a phase B' with similar properties. We believe that this behavior is a numerical artifact and that in fact B and B' are one and the same phase. 

\subsubsection{Spin-liquid candidate region (phase F)}

At the high-$\alpha$ side of the spiral phases we find an extended region where the spin-wave dispersion acquires imaginary modes. This means that MSW theory predicts an instability here. 
The width of this region in $\alpha$ stays approximately constant but it moves to smaller $\alpha$ with increasing temperature, leaving space to 
the collinear short-range-ordered phase (E). 
The region F extrapolates well down to the suspected spin-liquid phase between $1.35\lesssim\alpha\lesssim 1.66$ at $T=0$. 
Given that MSW is seen to break down at a putative spin-liquid phase at $T=0$ due to its lack of order, 
\emph{a fortiori} one can expect MSW to break down in the same parameter range at finite temperatures, 
because at finite $T$ the theory would be required to describe not only the ground state
but also the excitations on top of it. 

Note also that the spin-stiffness decreases upon approaching this region, which could be interpreted as a precursor of a short-range-ordered phase.

\subsubsection{2D-N\'{e}el states (phases D and E)}
As expected from BKT theory, when going to finite temperatures the 2D-N\'{e}el ground state first changes into a low-$T$ quasi-long-range ordered phase (phase D), which at a temperature $T_{\mathrm{BKT}}$ undergoes a transition into a high-$T$ short-range-ordered phase (phase E). 
Both are characterized by an ordering vector at the 2D-N\'{e}el value $\vect{Q}=\left(2 \pi,0\right)$. Furthermore neighboring spins which share a diagonal bond are strongly anticorrelated whereas neighboring spins which lie on the same chain are positively correlated.

The square XY lattice, which is reached as $\alpha\equiv t_2/t_1\to\infty$, has been extensively studied in the past. The classical BKT-temperature $T_{\mathrm{BKT}}^{\mathrm{cl}}/{t_2}=0.695$, which has been calculated by use of classical Monte Carlo simulations \cite{Cuccoli1995}, is significantly lowered in the quantum limit to around $T_{\mathrm{BKT}}/{t_2}\approx 0.35$ (Quantum Monte Carlo calculations \cite{Ding1990,Harada1997}).
Our MSW results yield a BKT temperature of $T_{\mathrm{BKT}}/{t_2}\approx 0.27$ at $\alpha=100$, where the system has practically reached the square lattice limit \footnote{At this value of $\alpha$ we computed the correlations to spins as far as 64 lattice sites away, contrarily to the rest of the phase diagram, see section \ref{cha:finiteT_observables}.}. 
Once again the disagreement with the MSW result is not surprising, given that this theory does not account properly for vortex--antivortex excitations. 
In particular the BKT line for $\alpha\gtrsim1.6$ is not very distinct, due to the location problems
mentioned in Sec.~\ref{cha:finiteT_observables}. 
Therefore its quantitative value should be interpreted with caution, especially in this parameter range. 
However, the qualitative behavior of the phase diagram seems to be described correctly. 

In the following section we turn to observables which show more clearly how order in the different phases persists at finite temperatures.

\subsection{Observables distinguishing between LRO and SRO \label{cha:observablesLRO-SRO}}

Here we focus on observables which help to distinguish between different types of regimes 
(\emph{i.e.\ }quasi-long-range order or short-range order), namely the partial Gaussian spin 
stiffness $\Upsilon^{\mathrm{partial}}$, the gap $\Delta$, the entropy, and the occupation of the zero mode $n_{\vect{k}=0}$.

\subsubsection{Entropy, spin stiffness, and gap}

The entropy [Eq.~\eqref{entropy}] shows behavior consistent with the quasi-ordered character of the low-temperature 2D-N\'{e}el phase D and the spiral phase B 
in the sense that it is smaller in phases with stronger order as can be seen in Fig.~\ref{fig:phd_finiteT_MSW}~(g). 
Correspondingly, in phases B and D $\Upsilon^{\mathrm{partial}}$ is large [Fig.~\ref{fig:phd_finiteT_MSW}~(d)] and $\Delta$ is very small [Fig.~\ref{fig:phd_finiteT_MSW}~(h)].

At the BKT transition no sharp change occurs in these observables as may be expected. 
However, the contour lines of the spin stiffness, the entropy, and the gap, all seem to be consistent 
with the shape of the $T_{BKT}$ curve.

We report the gap $\Delta$ for two representative values of $\alpha$ in Figs.~\ref{fig:Deltasn0s}~(a)~and~(b).
It evolves smoothly through the BKT transition in the Neel phase, whereas in the spiral phase it is seen to display a sharp increase right above the BKT transition up to the transition to phase A.
On the contrary, for all phases a sharp increase of $\Delta$ (accompanied by a sharp drop of $\Upsilon^{\mathrm{partial}}$) can be discerned at the breakdown temperature where correlations are completely lost.
    \begin{figure}
        \center
        \includegraphics[width=0.45\textwidth]{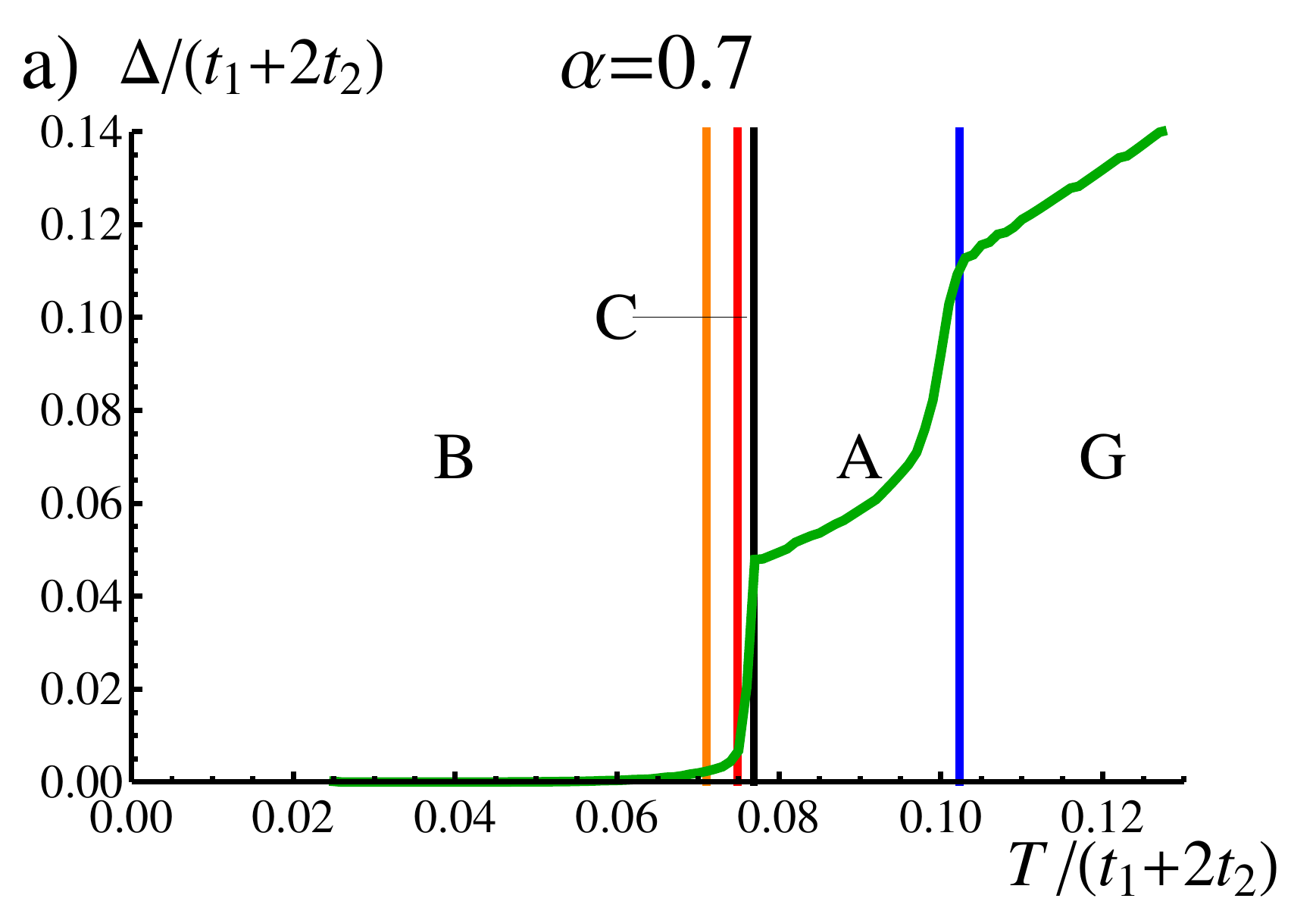}\quad\includegraphics[width=0.45\textwidth]{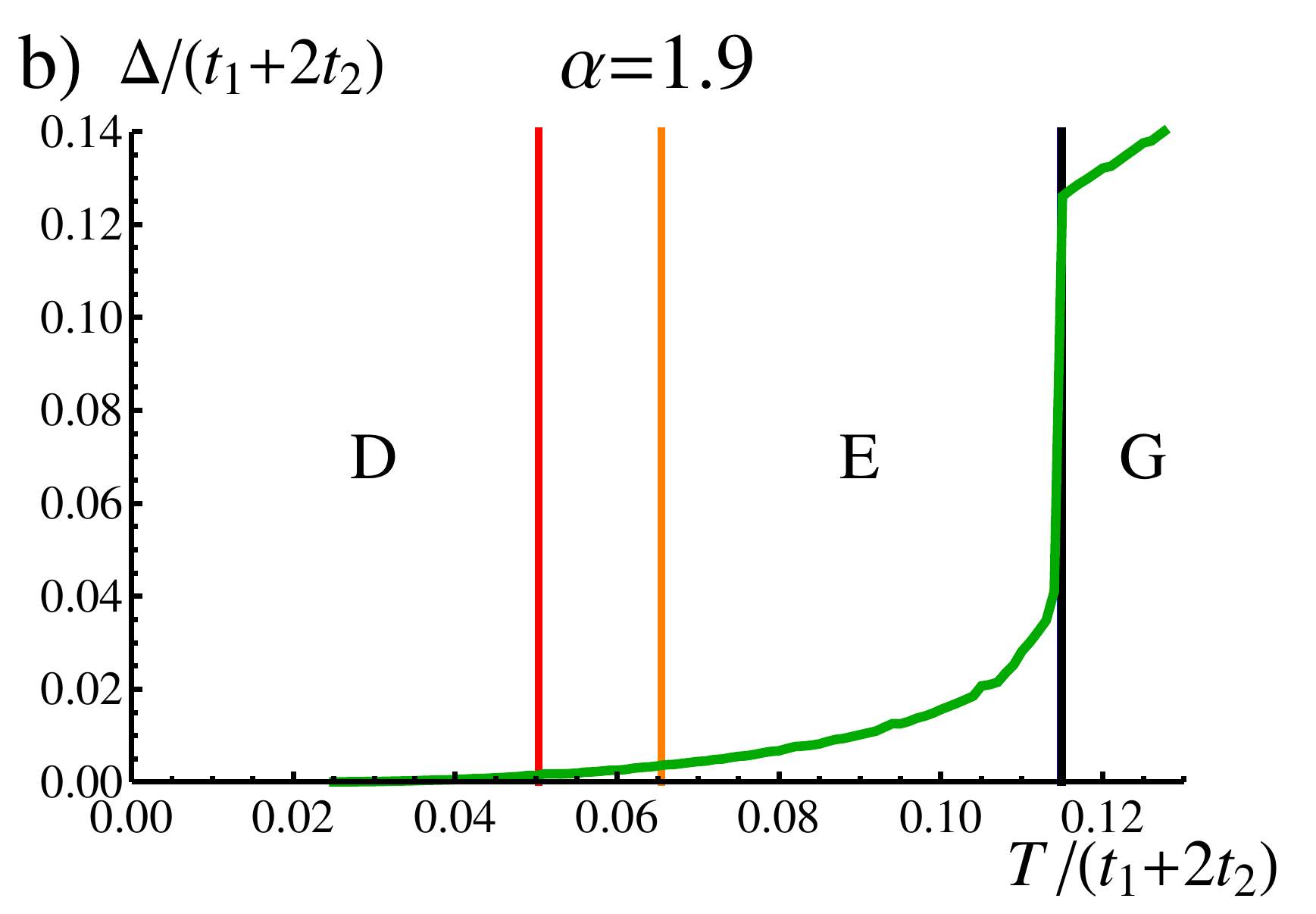}
        
        \includegraphics[width=0.45\textwidth]{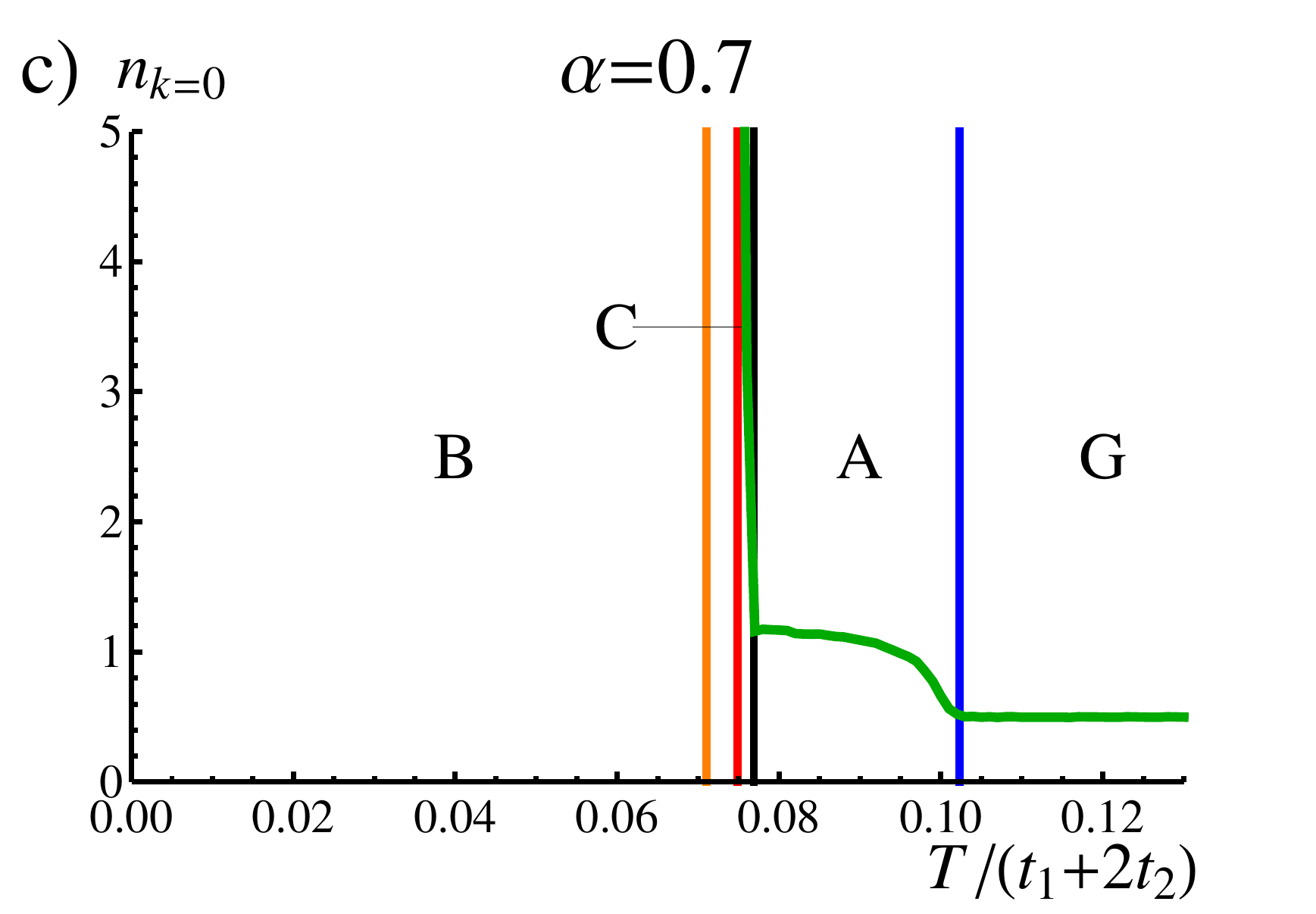}\quad\includegraphics[width=0.45\textwidth]{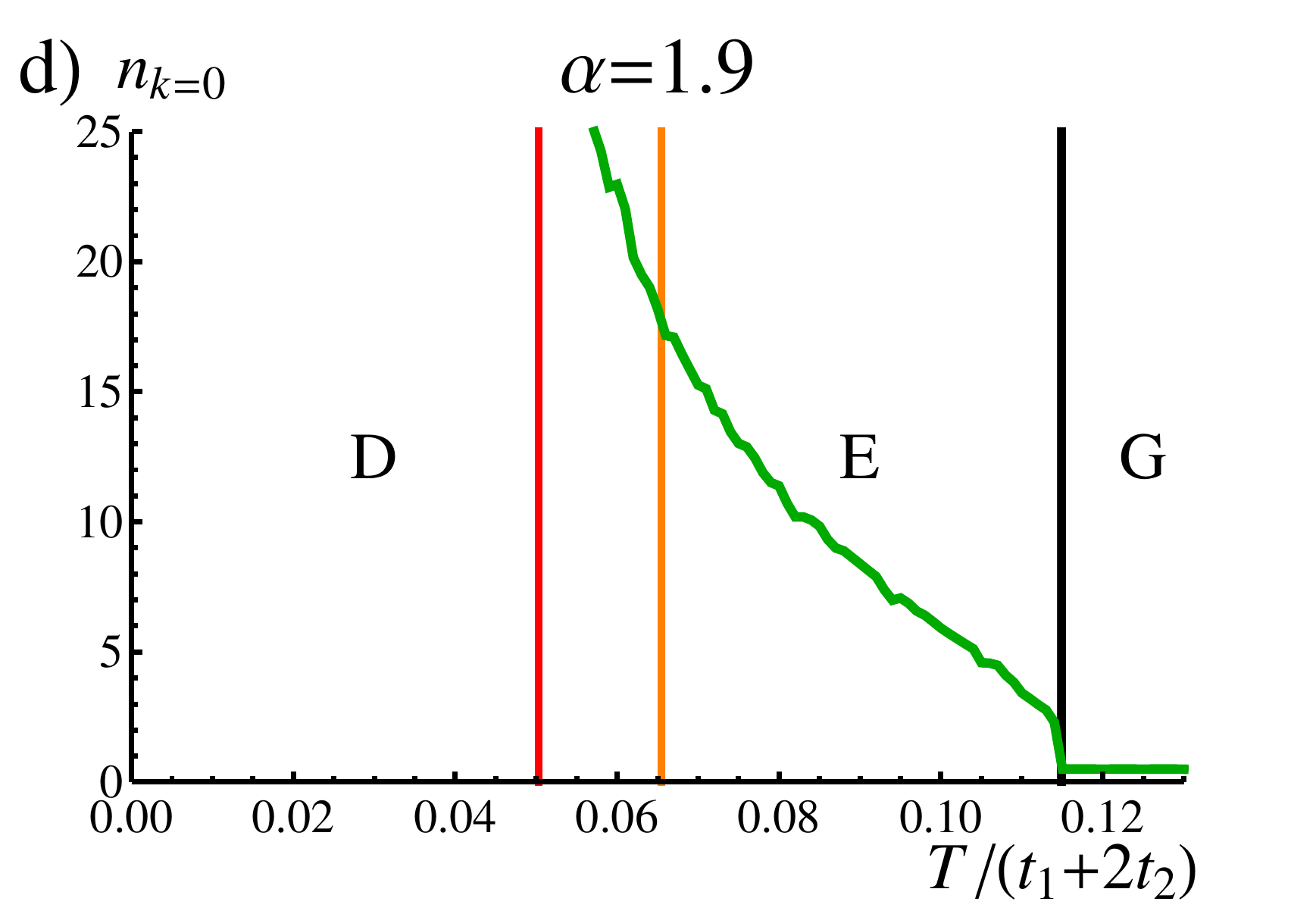}
        \caption{Gap $\Delta$ [(a) and (b)] and occupation of the zero mode $n_{\vect{k}=0}$ [(c) and (d)] for two values of $\alpha$. The vertical lines denote the critical temperatures of the BKT transition calculated via the decay of correlations in direction $\vs{\tau}_1=\left(1,0\right)$ (red), the corresponding BKT transition for $\vs{\tau}_2=\left(1/2,\sqrt{3}/2\right)$ (orange), the break-down temperature (blue), and the temperature where inter-chain correlations disappear (black) [in~(b) and~(d) the last two markers overlap], all computed through the two-point correlations $C_{m\, \vs{\tau}_{1,2}}$. Capital letters refer to the phases of Fig.~\ref{fig:TLXYfiniteTPhD}.}
        \label{fig:Deltasn0s}
    \end{figure}

In phase A (better visible for smaller values $\alpha$ which are not shown) we find that the gap is almost a linear function of temperature up to very close to the breakdown temperature. This is typical of critical systems where the temperature introduces the only energy scale, which is reflected in the gap.

\subsection{Occupation of the zero-mode}

The occupation of the zero-mode $n_{\vect{k}=0}$ [Figs.~\ref{fig:Deltasn0s}~(c)~and~(d)] gives an insightful measure of the strength of correlations. 
Due to constraint~\eqref{constraint} a large population of the zero mode entails smaller population of excited modes and therefore leads to stronger correlations.

Similar to what was seen for the gap in the previous section, at the transition between phases C and A $n_{\vect{k}=0}$ drops from very large values to something of the order of $1$. Afterwards it changes only slightly with $T$. It is useful to remember that the average mode occupation, $\sum_{\bm k} n_{\bm k}/N$,  
equals $S$ by virtue of constraint~\eqref{constraint}. This means that in the 1D-like short-range-ordered phase~A $n_{\vect{k}=0}$ is relatively small but still larger than the occupation of the other modes.

The behavior of  $n_{\vect{k}=0}$ is different for the 2D-N\'{e}el phase.
At the BKT line that we extracted from the analysis of the two-point correlation functions $C_{m\, \vs{\tau}_{1}}$ and $C_{m\, \vs{\tau}_{2}}$, $n_{\vect{k}=0}$ decreases strongly but smoothly.
Up to the breakdown point its values are still several times larger than in the 1D-like phase, however.
This supports our identification of the BKT transition; but the smoothness of $n_{\vect{k}=0}$ also shows the reason why the observables of section~\ref{cha:observablesLRO-SRO} could not point out a sharp transition.

\subsection{Discussion}

The finite-temperature phase diagram is observed to be a natural extension of the ground-state phase diagram.
We find that zero-temperature long-range ordered phases are reflected in finite-temperature phases with quasi-LRO, while phases with quasi-LRO at $T=0$ turn into short-range-ordered phases at any finite temperature.
When temperatures are at or below a few percent of the coupling strengths, the main characteristics of the ground state phase diagram are retrieved, with a short-range 1D-like phase (A), and two quasi-ordered phases [one with spiral properties near the isotropic triangular limit (B) and one with 2D-N\'{e}el-like characteristics at large values of $\alpha$ (D)] which are separated by a potential spin liquid (F). 
This last phase was identified by (i) the breakdown of MSW theory, which indicates that the assumption of an underlying ordered state is invalid, and (ii) the lowering of the spin stiffness as this phase is approached. In \ref{cha:appED} we show further evidence for spin-liquid behavior in this phase at T=0 gleaned from very limited exact-diagonalization results.

We have given a rough estimate for $T_{\mathrm{BKT}}$ over the entire range of anisotropies. We find agreement in the rough magnitude of $T_{\mathrm{BKT}}$ at points where estimations of the BKT temperatures computed by other methods exist. In our results the BKT transition is more clearly visible in the correlations for the spiral phase than for the 2D-N\'{e}el phase.

\end{section}


\begin{section}{Conclusions 		\label{cha:conclusion}}
We have extended Takahashi's modified spin-wave theory by an optimization of the ordering vector, which allows to account for order that deviates from the classical one.

We have used this method to calculate the ground state phase diagram of the spatially anisotropic triangular lattice with $S=1/2$ spins and XY interactions. We found the expansion of a quasi-ordered 1D-like phase to finite inter-chain couplings, a spiral phase, and a 2D-N\'{e}el phase. At the transition between the latter two 
the breakdown of MSW theory indicates the loss of LRO. 

We have extended this phase diagram to finite temperatures and computed Berezinskii--Kosterlitz--Thouless transitions, although the results are to be interpreted only semi-quantitatively because MSW theory does not explicitly account for vortex-antivortex excitations. 
 We find that the ground state phases clearly imprint their properties on the finite temperature phase diagram at low temperatures, with long-range ordered phases being replaced by quasi-ordered ones and quasi-ordered phases by short-range ordered ones.

Qualitative and even quantitative agreement with PEPS and ED calculations was found in the regions where magnetic LRO is to be expected. 
In particular it has been shown that MSW theory with ordering vector optimization is able to account satisfactorily for the main quantum corrections to the ordering vector. 
Furthermore, our calculations show that the spin stiffness is a useful observable for finding candidate regions for spin-liquid behavior in the ground state.
Indeed the breakdown of MSW theory, or the very weak stiffness of the magnetic order it predicts, can be used as indications of
the absence of long-range order in the exact ground state of the model. 

We find two main recurrent features for strongly frustrated quantum-magnets in two dimensions: collinear order is considerably stabilized 
by quantum and/ or thermal fluctuations against spiral order, and ordered or quasi-ordered phases characterized by different forms of order (collinear vs. spiral) 
do not continuously connect to each other, but they rather seem to be separated by quantum disordered phases. 
While MSW theory cannot determine the properties of such disordered phases, it provides a fast and clear method for \emph{finding} candidates of disordered phases. This method can therefore serve as a guide in our search for interesting quantum-mechanical lattice models which \emph{require} an experimental quantum simulator for further study of their phase diagram.

\end{section}


\begin{section}{Acknowledgments}
This work is financially supported by the Caixa Manresa, Spanish MEC/MINCIN project TOQATA (FIS2008-00784), EU Integrated Projects SCALA and AQUTE, and ERC Advanced Grant QUAGATUA.

\end{section}

\begin{appendix}
\begin{section}{Signatures of ordering and spin-liquid behavior in the exact diagonalization spectra of a small cluster \label{cha:appED}}

 We present here exact diagonalization data for the spectrum of the $S=1/2$ XY antiferromagnet 
 on the spatially anisotropic triangular lattice. We perform our calculations on a 24-spin cluster
 with the geometry depicted in Fig.~\ref{fig:systemsexactdiag}.
 The system Hamiltonian, Eq.~\eqref{HS}, commutes
 with the total magnetization along the $z$ axis, $S_z^{\mathrm{tot}}$, so that excited states can be classified
 on the basis of this quantum number \footnote{The momentum is not a good quantum number here, given that
 we consider open boundary conditions.}.   
 
 \begin{figure}
        \centering
        \mbox{
        \includegraphics[width=0.33\textwidth]{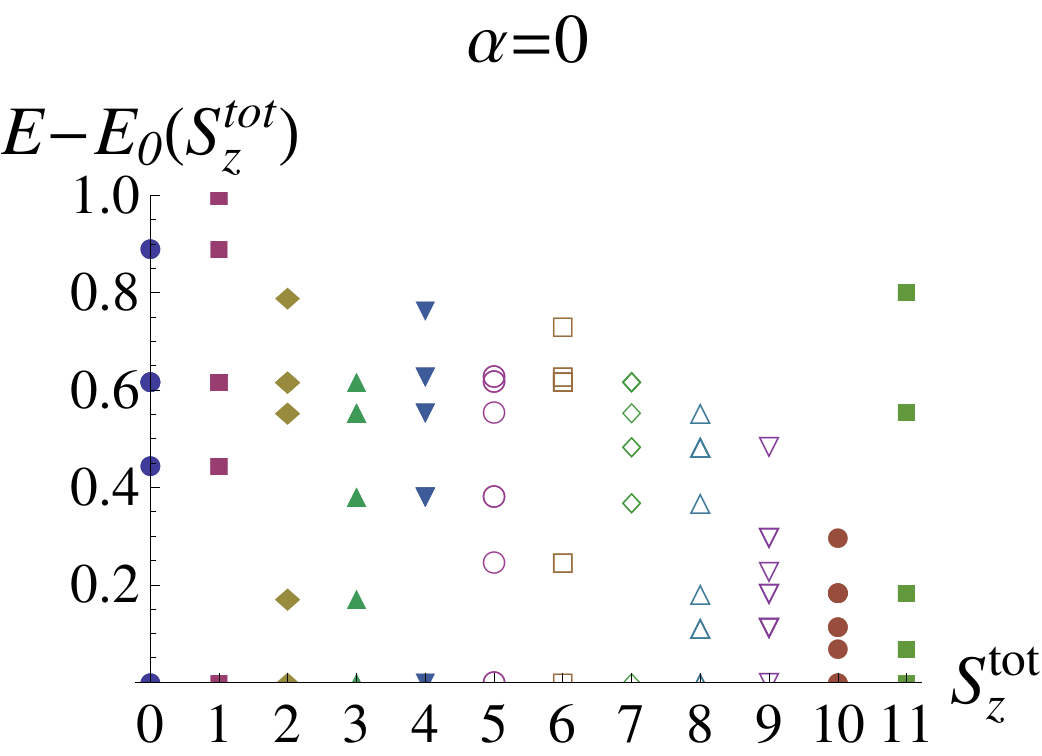}
        \includegraphics[width=0.33\textwidth]{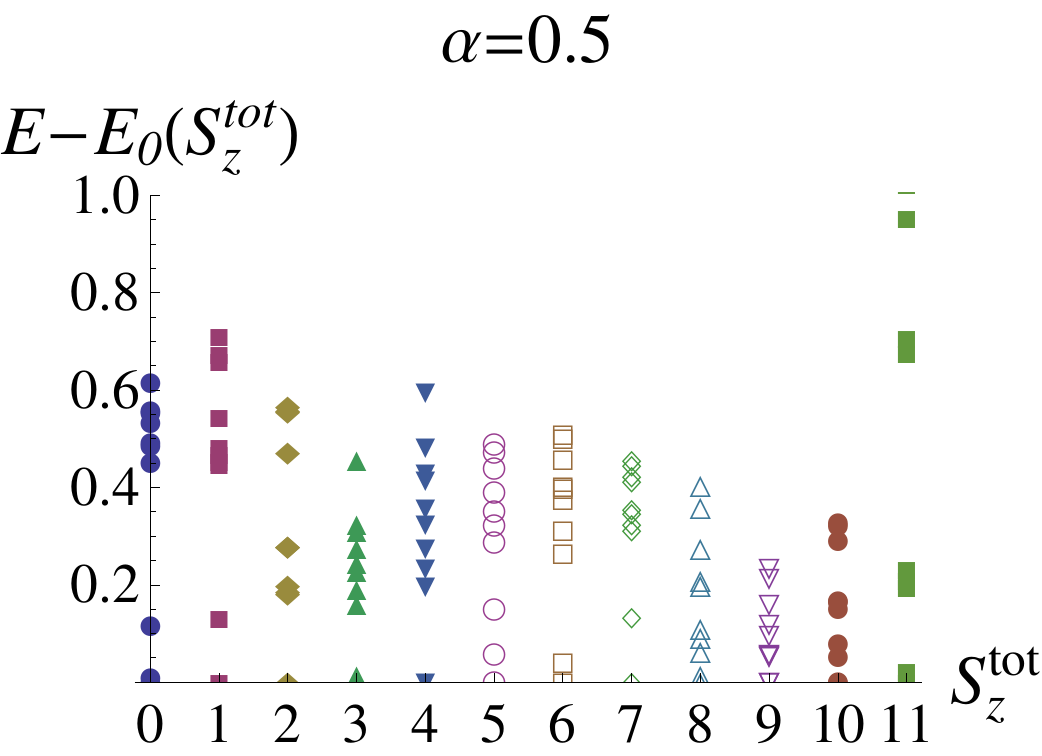}
        \includegraphics[width=0.33\textwidth]{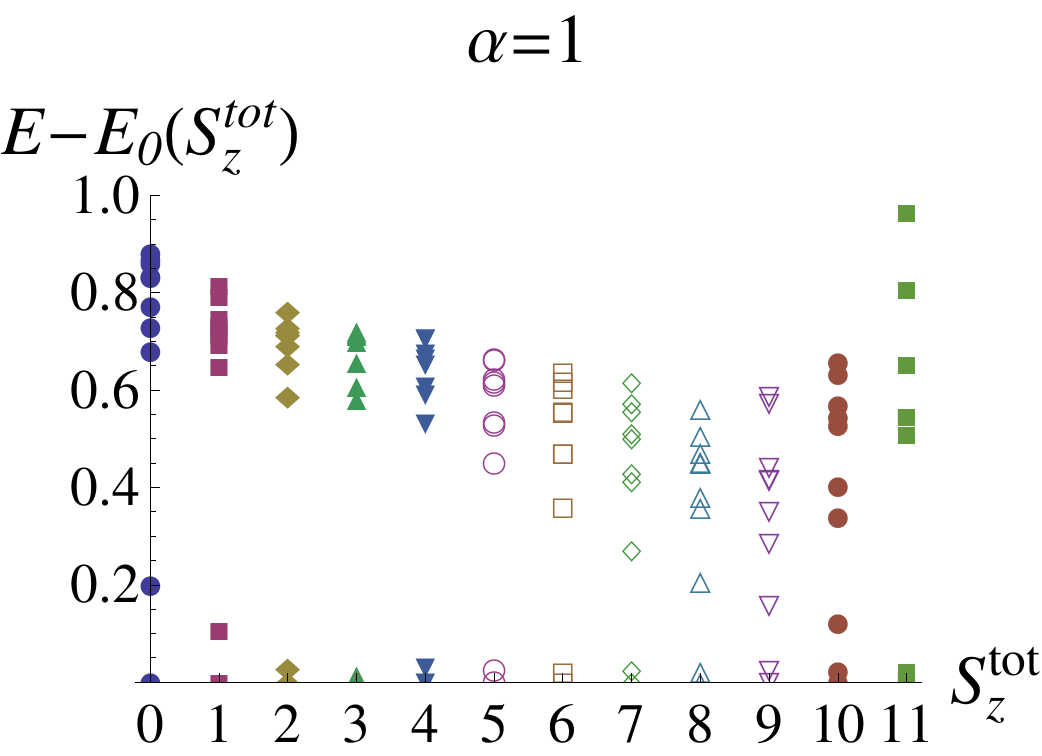}} \\
        \hspace{0.5cm}
        \mbox{
        \includegraphics[width=0.33\textwidth]{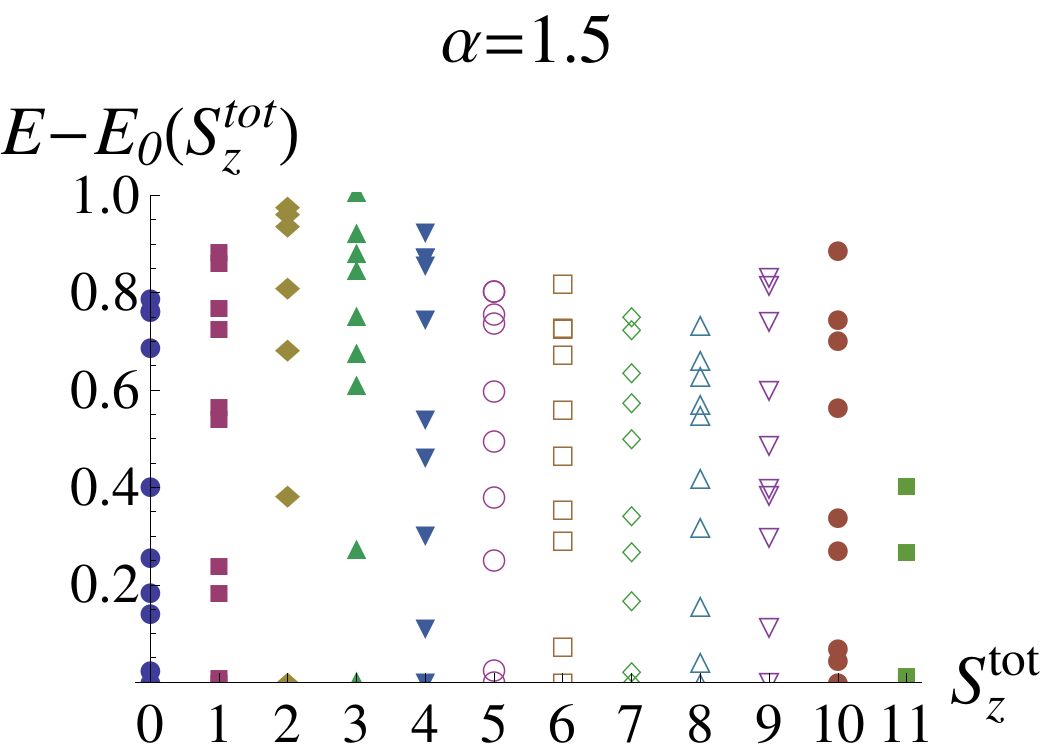}
        \includegraphics[width=0.33\textwidth]{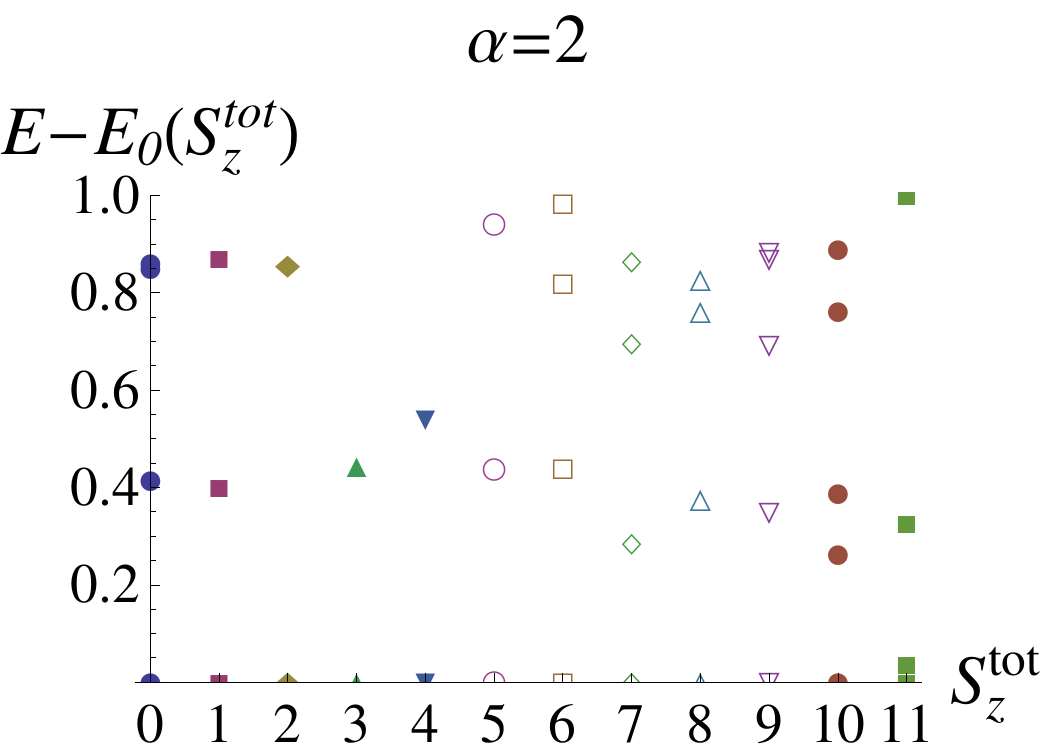}
        \includegraphics[width=0.33\textwidth]{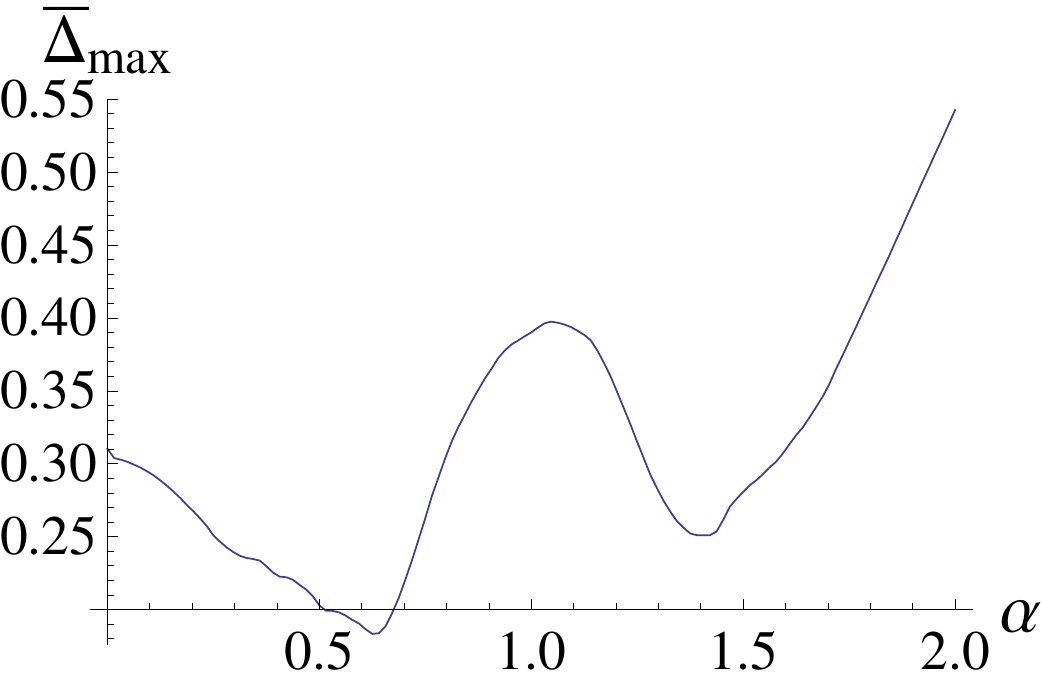}}
        \caption{Spectra of the $S=1/2$ XY antiferromagnet 
 on the spatially anisotropic triangular lattice from exact diagonalization on a 24-site cluster, 
 for various values of the lattice anisotropy $\alpha$.
 The lower-right panel shows the average maximal level spacing (see text)
  as a function of $\alpha$.}
        \label{fig:spectrum}
    \end{figure}

Figure~\ref{fig:spectrum} shows the excitation energies of the first excited states
in each $S_z^{\mathrm{tot}}$ sector (up to $S_z^{\mathrm{tot}}=11$) with respect to the minimum
energy in each sector, $E_0(S_z^{\mathrm{tot}})$ [$E_0(S_z^{\mathrm{tot}}=0)$ corresponds to 
the ground state energy]. Upon varying $\alpha$ we observe a significant 
evolution of the low-energy spectrum of the system, which points at 
the widely different regimes explored by the system. In particular, in the 
spirally and N\'eel ordered phases -- exemplified in Fig.~\ref{fig:spectrum}
by $\alpha=1$ and $\alpha = 2$, respectively -- we observe that 
in each $S_z^{\mathrm{tot}}$ sector there are a few states lying close to the 
minimum energy one, and separated from the other excited states
by a large gap. According to a standard `tower-of-states' argument
\cite{Lhuillier2005}, these low-lying states
are expected to collapse to the ground state in the thermodynamic limit, 
giving rise to degenerate superpositions of all $S_z^{\mathrm{tot}}$ sectors, each breaking
the U(1) rotational symmetry of the Hamiltonian and displaying 
spiral or N\'eel order. The higher-energy states will instead reproduce the 
true excitation spectrum in the thermodynamic limit. 

 This tower-of-states feature is on the contrary absent in other regions
 of the phase diagram, in which the energy levels in each $S_z^{\mathrm{tot}}$
 sector are more homogeneously spaced. The absence of a low-lying
 multiplet of states separated from the higher energy states by a large
 gap is observed in models whose ground state is generally
 considered to be a spin liquid \cite{Lecheminant1997}. We therefore introduce an observable aimed at
quantifying the extent to which the spectrum exhibits the expected
features in presence of spontaneous symmetry breaking
in the thermodynamic limit. We consider the \emph{average maximal
level spacing} $\bar{\Delta}_{max}$, defined as
\begin{equation}
\bar{\Delta}_{max} = \frac{1}{N_S+1} \sum_{S_z^{\mathrm{tot}}=0}^{N_S} 
\max_{i} \left[ E_{i+1}(S_z^{\mathrm{tot}}) - E_{i}(S_z^{\mathrm{tot}}) \right]~,
\end{equation}
namely $\bar{\Delta}_{max}$ is the maximal level spacing in each 
$S_z^{\mathrm{tot}}$ sector, averaged over the $N_S+1=12$ sectors considered.
The maximal level spacing is extracted by considering the lowest $10$ levels $E_{i}(S_z^{\mathrm{tot}})$, which captures the behavior of the low-energy part of the spectrum.
The above quantity is chosen so that it will be maximal in presence of
a large separation between the low-lying tower of states and the higher-energy 
spectrum, while it will be minimal in presence of homogeneously
spaced levels in each sector. 

 When plotting $\bar{\Delta}_{max}$ as a function of 
 $\alpha$, as shown in Fig.~\ref{fig:spectrum}, we observe two 
 very pronounced relative minima, at $\alpha\approx  0.6$
and $\alpha\approx 1.4$. Remarkably these two minima correspond
to the two regions in parameter space where PEPS calculations
predict the occurrence of a spin-liquid phase \cite{Schmied2008}
(compare Fig.~\ref{fig:phasediagtriang}). Hence the lack of the
tower-of-states feature in the spectra of a small cluster is
consistent with the PEPS prediction. 

\end{section}

\end{appendix}


\vspace{1cm}
\bibliographystyle{nature}
\bibliography{D:/work/Diplom/thesis/references}

\end{document}